\documentclass[manuscript=article]{achemso}

\setkeys{acs}{etalmode = truncate}
\setkeys{acs}{maxauthors = 10}

\usepackage{subcaption} 
\usepackage{stmaryrd} 
\usepackage{amssymb}
\usepackage{amsmath}

\usepackage{tabularx}
\newcolumntype{Y}{>{\centering\arraybackslash}X} 
\usepackage{multirow}
\usepackage{booktabs} 

\usepackage{xcolor}

\author{Johannes L. Teunissen}
\affiliation[BIRA]{Royal Belgian Institute for Space Aeronomy, Av Circulaire 3, 1180 Brussels,
Belgium.}
\alsoaffiliation{Univ. Lille, CNRS UMR 8520 - IEMN - Institute of Electronics, Microelectronics and Nanotechnology, F-59000 Lille, France}
\email{jlteunissen@gmail.com}

\author{Fabiana Da Pieve}
\affiliation[BIRA]{Royal Belgian Institute for Space Aeronomy, Av Circulaire 3, 1180 Brussels,
Belgium.}

\title{Molecular Bond Engineering and Feature Learning for the Design of Hybrid Organic-Inorganic Perovskites Solar Cells with Strong Non-Covalent Halogen-Cation Interactions}

\begin{document}



\begin{abstract}
Hybrid organic-inorganic perovskites are exceedingly interesting candidates for new solar energy technologies, for both ground-based and space applications.
However, their large-scale production is hampered by the lack of long-term stability, mostly associated to ion migration. The specific role of non-covalent bonds in contributing to the stability remains elusive, and in certain cases controversial.
Here, we perform an investigation on a large perovskite chemical space via a combination of first-principles calculations for the bond strengths and the recently developed Sure Independent Screening and Sparsifying Operator (SISSO) algorithm. The latter is used to formulate mathematical descriptors that, by highlighting the importance of specific non-covalent molecular bonds, can guide the design of perovskites with suppressed ion migration.
The results unveil the distinct nature of different non-covalent interactions, with remarkable differences compared to previous arguments and interpretations in the literature 
on the basis of smaller chemical spaces. In particular, we clarify the origin of the higher stability offered by FA compared to MA, which shows to be different from previous arguments in the literature, the reasons of the improved stability given by the halogen F, and explain the exceptional case of overall stronger bonds for Guanidiunium. 
The found descriptors reveal 
the criteria that, within the stability boundaries given by the Goldschmidt factor, give an all-in-one picture of non-covalent interactions which provide the more stable configurations, also including interactions other than H-bonds. Such descriptors are more informative than previously used quantities and can be used as universal input to better inform new machine learning studies.
\end{abstract}

\section{Introduction}

Hybrid Organic-Inorganic Perovskites (HOIPs)-based solar cells are emerging as a new opportunity for a paradigm shift in solar cell technology for both Earth-based\cite{Jeon2015} and space applications\cite{brown,Cardinaletti2018,yangspace}.
On Earth, the increasing demand for clean energy resources, rapidly evolving high power conversion efficiency (PCE)\cite{MaxPCE, efficiencytables} of HOIPs and the relatively low-cost fabrication \cite{Snaith2013} make them potential candidates for a step-change approach to harvest solar energy.
For space missions, HOIPs qualify as the best candidates for extremely lightweight\cite{Cardinaletti2018}, flexible and radiation resistant solar panels.\cite{Kaltenbrunner2015, Burschka2013Flexible, Lang2016, Miyazawa2015, Miyazawa2018}
Either for missions in the Earth's vicinity or deep Space, these systems show potential for reducing the weight at launch, for broad absorption that matches the operating conditions at both Earth and Martian orbit and for  fabrication via 3D printing techniques, considered as necessary for on-mission manufacturing\cite{Huang2017, Cardinaletti2018}. For stays on the Moon or Mars, these systems offer the potential for autonomous roll-out of solar arrays for extra-terrestrial surface deployments. 


HOIPs comprise a negatively charged lead-halide inorganic skeleton, with a composition of ABX3, where B is a metal cation (Sn2+ or Pb2+), X is a halide anion (I-, Br-, and/or Cl-) and A is a monovalent positively charged organic cation, such as methylammonium (MA=CH3NH3X+ where X=I, Br, Cl) or formamidinium (FA=CH(NH2)2+).

The organic cation plays a fundamental role for the stability of the structure. Perovskite materials have excellent optical and electrical properties for outstanding performance in solar cell applications, such as strong light absorption \cite{Kojima2009, YUE2016},
steep optical absorption edge \cite{DeWolf2014}, band gap tuning \cite{mcmee},
high charge carrier mobility \cite{Saba2016, Motta2015,Wehrenfennig2014}, 
possible polaron-protected transport properties\cite{zheng2019,frost2017_2,ambrosio2018} 
(although debated\cite{schlipf}), low exciton binding energy \cite{Miyata2015, Kandada2016}
and long charge carrier diffusion lengths and carrier lifetimes.\cite{Ponseca2014, Stranks341, Wehrenfennig2014}

The recent steep increase in PCE of perovskite solar cells 
makes them the fastest-advancing solar technology.
Nevertheless, despite the many advantages, the progress in large scale applications and production of HOIPs is hampered by instability issues.
Indeed, HOIPs suffer from structural instabilities induced by air, moisture\cite{boyd}, UV light\cite{berry,chinphyslett,akbulatov}, heat,  
causing the perovskite solar cells to have a particularly low lifetime. Intrinsic instability causing unavoidable decomposition with time has also been reported\cite{chinphyslett}. 
The instability is mainly caused by a relatively weak cohesion between the organic cation and the inorganic octahedra, and predominantly by the low energy barriers for the 
migration of halide anions and organic cations, with halide migration being the most prevalent.
\cite{Eames2015, Meggiolaro2018, oranskaia2019, Futscher2019, Khan2020}. Nevertheless, the (short-term) performance of HOIPs-based solar cells is considerably defect tolerant. Density Functional Theory (DFT) calculations based on hybrid functionals with spin orbit coupling have shown that the majority of the defects induce only shallow defect states near the band edges\cite{adinolfi,yin2014}, and that although iodine interstitials create deep levels in the gap of MAPbI$_3$\cite{du2015,meggio2018ACS}, these only leave  short-living hole traps\cite{Meggiolaro2018,WOLF2020217}. 
Still, there remain unwanted effects in carrier dynamics, I-V hysteresis and in the chemical degradation of the very HOIP and at the interfaces with the charge transport layers. Ion migration has also been reported to greatly accelerates charge carrier losses\cite{prezhdo} and, in large scale, also to cause phase segregation.\cite{Lang2018}

A tremendous experimental effort has been devoted to overcome the above shortcomings of HOIPs, though intrinsic improvements (mostly by changing the stoichiometry) or extrinsic improvements (such as encapsulation) that reduce exposure to degradation factors. For example, intrinsic improvements consist in changing the amount of or mixing different organic cations, such as
through insertion of excess or combined MA/FA,\cite{Yang2016,YiLuoMeloni2016,Ferdani2019}
introducing small amounts of larger organic molecules\cite{wang}, inserting guanidinium\cite{DeMarco2016,Ferdani2019}, or changing the amount of iodine\cite{Yang1376} and or mixing different halides\cite{noh}, via mixed-cation systems (for example by including inorganic cations such as rubidium or caesium)\cite{grat}, through doping for lattice strain relaxation-induced suppression of halides vacancies,\cite{Saidaminov2018}
multi-vacancy defects passivation, \cite{Zheng2017} through lower-dimensional HOIPs or additional layers,\cite{LiLiu2017},
and chemical bond modulation tuning the bond strength of the additives.\cite{LiTao2019, HolekeviC2021}

Halogen migration can be mitigated by increasing the binding strength with the organic cations.\cite{Yang1376, ZhangTeng,oranskaia2018}. The organic cation binds strongly with the halogen anions via hydrogen bonding, and when the cation has double bonds, also via $\pi\cdots$anion binding.\cite{anionpi}
Several papers have shown that in MAPbI$_3$ only the $N\!-\!H\!\cdots\!I$ hydrogen bonds are important\cite{Lee2015, Varadwaj2019, Lee2016}, although recently it was advocated\cite{Varadwaj2019} that non-covalent interactions other than $N\!-\!H\!\cdots\!I$, such as $C\!-\!H\!\cdots\!I$ and $N/C\!\cdots\!I$  are also of importance.
Nonetheless, neither computational nor experimental methods can easily deconvolute the individual contributions to the energetics of the bonds, as a continuous merge between the hydrogen bond with van der Waals interactions and ionic and covalent bonds\cite{Svane2017}, and despite the tremendous efforts, the unique properties of metal halides perovskites remain elusive.

In this work, we combine a first-principles analysis of bonds based on the Quantum-Theory-of-Atoms-In-Molecules (QTAIM)\cite{bader1990} with a feature learning analysis through the Sure Independence Screening and Sparsifying Operator (SISSO) algorithm\cite{Ouyang2018} on a large chemical space, 
to untangle the contribution from different bond types and investigate the features that dominate in the relationship between the structural composition of the HOIPs and the X-cation binding strengths.
The findings provide new insights into the role of different non-covalent interactions, with considerably different interpretations to previous arguments in the literature on the basis of some few specific cases, and can be used to guide the design of more stable cells by carefully tuning chemical and compositional/structural properties for ion migration suppression.

\section{\label{sec2}Theory and Computational details}

\subsection{QTAIM and DFT Calculations}

In order to find the binding strengths, we combine the "Atoms in Molecules" theory\cite{bader1990} and the empirically established relationship between the properties of the BCPs\cite{MATA,oranskaia2019}.
Within this theory, a bonding interaction is defined by a maximum electron density path (bond path) connecting two interacting atoms.
Critical points (CPs) are points where the gradient of the electron density, $\nabla \rho$ becomes zero. 
Bond critical points (BCPs) are those type of critical points where the density increases in one direction while it decreases in the plane perpendicular to that direction. 
Thus, a BCP is a saddle point in the charge density along the bond path. The existence of BCP between the acceptor atom and the donating hydrogen, as well as the charge density properties at this point, defined the major criteria of the existence of an H-bond.
These BCPs are often found close to the midpoints of two atoms that have a bonding interaction, either covalent or non-covalent. 
The estimation of the bonding strength starts from the Abramov's formula\cite{Abramov} for the local kinetic energy density, $G$, given as:
\begin{equation}
G = \frac{3}{10} (3\pi^2)^{\frac{2}{3}} \rho^{\frac{5}{3}}
+ \frac{1}{72} {\left | \nabla \rho \right |}^{-1}
+ \frac{1}{6} \nabla^2 \rho
\label{abramov}
\end{equation}
where $\rho$ is the electron density.
$G$ correlates linearly with the bonding strength.\cite{MATA}
At the BCPs, the second term vanished since the gradient is zero.
The presence of a BCP indicates there is a charge accumulation between the two bonded nuclei.\cite{KUMAR2016} 
The larger the density at the BCP, the stronger the interaction.
Negative gradients indicate covalent bonds while non-covalent interactions show small but positive gradients.\cite{KUMAR2016}

Mata \latin{et al.} \cite{MATA} fitted the bond strengths vs. $G$ and 
found a relationship with the bond strengths $E$ as: $E=0.429G$ with a high correlation of $R^2=0.990$.
The absolute values obtained for the bond strengths do depend on the level of theory \cite{theorydependenceAIM} but the interest of this work lies predominantly in the trends, hence, only relative differences matter.
There exist some criticism concerning the bond critical points\cite{shant2018}, as some cases exists where there is a bonding interaction without a BCP being present, most notably 1,2-ethanediol \cite{Lane2013}, but these cases seem to occur seldom and in our study we did not find any absence of BCPs for cation-halogen bonding interactions so we reasonably expect our trends to hold.




For all the non-covalent interactions analyzed in this study between an organic cation and a halogen, there is a clear and strong power law correlation between the bond lengths and the density at the bond critical point. For an analysis of the density and gradients at the BCPs, see Supporting Information Section 4.

%
%
As our starting set of perovskite structures, we take the dataset as prepared and provided by Kim \latin{et al.}\cite{Kim2017}\ of 1346 hybrid organic-inorganic perovskites. 
From this dataset we only select those 548 that have a perovskite structure i.e. roughly a cubic arrangement of the metal atoms with the halogens surrounding these metals connected via corner sharing octahedrals and the organic cation in the cavity thus created.
Finally, per chemical composition, we only take the most stable one, i.e. the one with the lowest atomization energy (the energy required to separate the molecular structure into individual atoms) resulting in a final set of 190 HOIPs. 
For some chemical compositions the perovskite structure is not the most stable structure. However, in this paper we are primarily interested in the binding strength {\it under the condition} that the structure is perovskite.
When the structure is non-perovskite, the binding characteristics are completely different, and we cannot compare the binding strengths between different chemical compositions.
For example for many hydroxylammonium or hydrazinium compounds, (like Hydroxylammonium Germanium Fluoride) the most stable structure is not a perovskite.
Although for a certain chemical composition, the perovskite structure might not be the most stable, it might still provide relevant insight in how to design structurally stable perovskites.
Moreover, in real applications one often uses a mixture of different organic cations,
(for example MA$_x$FA$_{(1-x)}$PbI$_3$) and for the mixed compound the perovskite structure might well be stable\cite{Ava2019}. 

For each of the 190 HOIPs, the BCPs between the organic cation and the halogen are determined.
In every perovskite structure there are three symmetrically unique halogen positions. 
Each inorganic cavity, which shape is approximately cubic (most generally a paralellepiped), has 12 halogens close to the midpoints of the cube edges and thus located approximately in the [002], [020] and [200] planes of the cavity.
Each organic cation located in the approximately cubic cavity, can bind maximally to 12 halogens, of which there are 3 sets of 4 halogens that belong to the same unique crystallographic position. 
We are mainly interested in those of the three halogens that has the lowest cation binding energy, since the weakest bound halogen will be most prone to halogen migration.
In this paper we will call the total bond strength between a single organic cation and its neighboring halogens $E_{total}$. This bond strength is the sum of the contribution from the 3 unique halogen positions. The bond strength of the halogen with the lowest cation binding energy is called $E_{minX}$.

%
The HOIP dataset as prepared by Kim \latin{et al.}\ uses 3 metals (Ge, Sn, Pb), 4 halogens (F, Cl, Br, I) and 16 organic cations giving 192 ($3\times4\times16$) unique chemical compositions.
For each composition multiple crystallographic structural conformations are present.
In total 1346 structures were obtained using a minima-hopping method\cite{minimahop} followed by a DFT-PAW structure optimization including van der Waals dispersion interactions via the vdW-DF2 functional\cite{vdwDF2}, using VASP \cite{VASP1}.
This dataset is open-source and can also be found on the NoMaD database.\cite{NOMAD}
The crystallographic information format (cif) files of the dataset where read using the \texttt{pymatgen} library.\cite{pymatgen}. 

For the systems with a perovskite structure, we calculated the continuous symmetry measures for cubicity\cite{CSMPinsky} of the metal atoms and found that for most of the HOIPs the metals are arranged in an approximately cubic shape.
The HOIPs with the highest deviations from cubic cells contained propylammonium or butylammonium cations (see Electronic Supplementary Information).
For two chemical compositions, no perovskite structures are present in the dataset, being ammonium lead fluoride and hydroxylammonium tin bromide. Hence, in this paper we study 190 instead of 192 structures. 
All the electron densities were obtained with Quantum ESPRESSO 6.3\cite{QE2} using the HSE06 functional\cite{HSE} and norm-conserving scalar relativistic pseudopotentials.\cite{HSE}
The kinetic energy cutoff of the wavefunction was set at 80 Ryd and the cutoff of the charge density to 320 Ryd.
The HSE06 functional is regularly used for HOIPs and is known to give good electronic and structural results although band gaps are slightly overestimated.\cite{Kim2017, oranskaia2019, Yu_2019}
A $5\times5\times5$ Monkhorst-Pack grid was used, centered on the $\Gamma$ point. 
The electron densities were evaluated at a $400\times400\times400$ grid and the critical points were obtained using the Critic2 program.\cite{Critic2}
The bond strengths presented in this work are based on the structures with the lowest atomization energies. However, HOIPs are particularly known to undergo subtle phase transitions.
At lower temperatures some symmetry breaking occurs changing the structure from cubic to tetragonal to orthorombic. 
However, the cation-halogen interactions are not expected to be significantly different, as they are not for MA, one of the most mobile cations.\cite{VARADWAJ2018}
Due to the corresponding differences in crystal packing, the orientation and bond lengths of the organic cation with the halogens will change. Nevertheless, the nature of the bonds and the ability of the organic cation to bind with the halogen anions will stay the same, and the observed trends stay the same as long as the structure is a perovskite 
(For a comparison of all 548 perovskite structures see Supporting Information Section 2).
Also Varadwaj \latin{et al.}\cite{Varadwaj2019}\ found that the binding characteristics stay the same for different cation orientations although Lee \latin{et al.}\cite{Lee2016}\ found that the energy difference between two different orientations of the MA cation in MAPbI$_3$ might be 45 meV. Since we found that the total bond strength for MAPbI$_3$ is 413 meV, this means that the binding strength might change by around 10\% depending on the cation orientation.
Taking this all in consideration, we believe that the results presented in this paper show valuable trends in the binding characteristics of different perovskite structures that can be extrapolated to other crystal packings, for example those of mixed cation or mixed halogen HOIPs.

\subsection{Sure Independence Screening and Sparsifying Operator (SISSO) algorithm}

Following the calculation of all $E_{total}$ and $E_{minX}$, we try to establish a connection between the bond strengths and the chemical structure via symbolic regression.
Symbolic regression tries to find a mathematical expression of the input features that best fits the data.
Here we try to find an expression of structural features that best predicts the cation-halogen binding strengths.
The particular algorithm used is called SISSO which is based on a compressed sensing technique (\textit{vide infra})\cite{Ouyang2018, Ouyang2019,ExtendTol,XIE2020}.
It starts from a set of single-valued elementary features, called $\Phi_0$. It then recursively applies some unary and binary mathematical operations to create new feature spaces called $\Phi_i$ with $i$ being the number of times the mathematical operations are applied. It then iteratively applies the Sure Independence Screening (SIS) and the Sparsifying Operator (SO) until the desired dimension is reached.
SIS selects a subspace of $\Phi_i$ that has the highest correlation with the target property (in our case the binding strengths) and SO selects the best descriptor from the union of all the subspaces selected by all the previous SIS steps.\cite{Ouyang2018}

In our case, we used SISSO to extract the most important structural features that dictate the cation-X binding strengths, thus elucidating what kind of structural composition is optimal for HOIPs with suppressed X-migration.
%
The included unary operations are 
($\boxempty^2,\sqrt{\boxempty}, \boxempty^{-1}, \left|\boxempty\right|$)
and the included binary operators are ($+,-,*,/$).
We also set a maximum to the descriptor complexity thus limiting the number of mathematical operations.
This prevents the loss of scientific interpretability as the higher complexity descriptors become too convoluted and from a certain point higher complexity features do not give significantly better results and can be seen as a measure to prevent over-fitting.
Setting the maximum complexity can be seen as a measure to prevent over-fitting.

We decided to only include features that can be derived directly from the given structure. No features (such as bond lengths or band gaps) are included that would require the knowledge of the specific perovskite structure. This enables direct evaluation of the descriptors, as is also possible for the Goldschmidt tolerance factor. 
The included features are the effective crystal radii of the metal, halogen and organic cation\cite{Kieslich2014} as well as the electronegativity and polarizability of the metal and halogen atoms.
The latter two features have shown to be useful in multiple studies combining HOIPs and predictive techniques.\cite{Lu2018, XIE2020} 
Differently from previous studies and in order to easily encode the nature of the organic cations, here some additional bond counts were included: the number of C-H, N-H and O-H bonds as well as the number of $\pi$ bonds. 
Hydroxylammonium is the only cation in the dataset that contains an O-H bond.
There are 4 organic cations that contain $\pi$ bonds: Formamidinium, Acetamidinium, Guanidinium and Imidazolium with respectively 2, 3, 3 and 5 $\pi$-bonds.
Evidently, there is a correlation between the electronegativity, $\chi$, and the polarizability, $\alpha$, of the halogens and metal, however the SISSO algorithm does not suffer from highly correlated features \cite{Ouyang2018}.

\section{\label{sec3}Results and Discussion}

\subsection{Connecting stability of HOIPs with composition and structure}


In this section we will assess the non-covalent binding interactions between the organic cation and the halogen atoms in a single unit cell consisting of one (approximate) cube formed by the metal atoms, plus the 12 halogens along the edges of the cube and the single organic cation inside the cube.
As stated before, we consider the total binding strength of all cation-X interactions called $E_{total}$ and the binding strength corresponding with the halogen that has the weakest interaction with the cation called $E_{minX}$. Since there are three different halogens, $E_{minX}\leq E_{total}/3$.

\begin{figure}[htbp]
\centering
\noindent\includegraphics[width=0.48\textwidth]{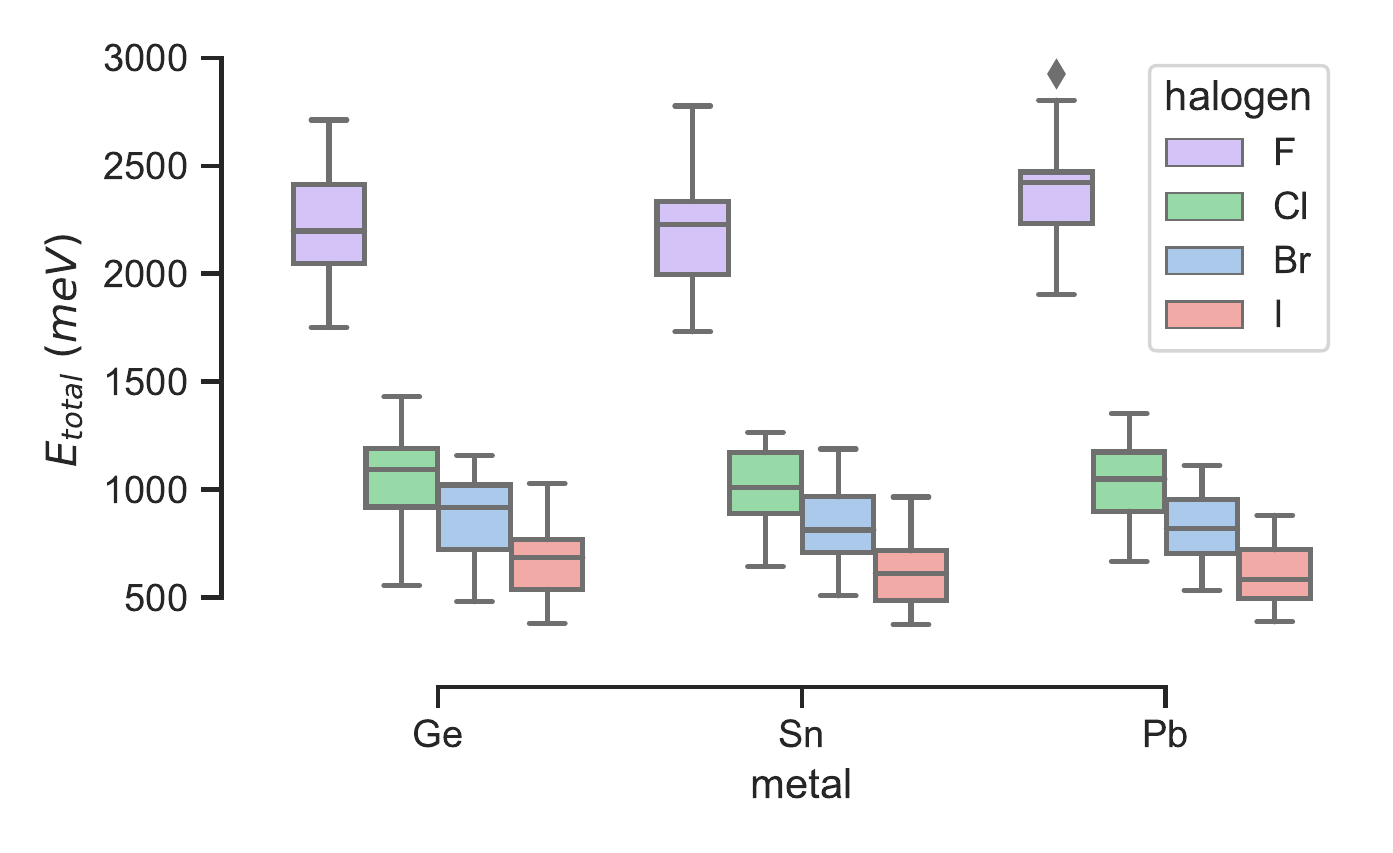}
\caption{A boxplot showing the distribution of the $E_{total}$ values grouped by metal and halogen. The boxes show the quartiles of the dataset while the whiskers are extended to 1.5 times the interquartile range (IQR).}
\label{fboxplot}
\end{figure}

The distribution of all the $E_{total}$ values obtained for all the 190 HOIPs is presented in Figure \ref{fboxplot}.
The type of halogen atom most clearly affects $E_{total}$.
With fluorine, $E_{total}$ is about twice as high as for heavier halogens. 
This trend is not only found in HOIPs\cite{elmellouhi} but is typically for any $D\!-\!H\!\cdots\!X\!\!-\!M$ interaction where $D$ is either C, N, or O.\cite{Brammer2001}
The high electronegativity of fluorine causes the $X\!\!-\!M$ bonds to be of considerably polar character leaving a strong negatively electrostatic potential around the atom, causing strong hydrogen bonds, generally called metal-assisted hydrogen bonds.\cite{kovacs2006}
Changing the metal has only a minor influence, in agreement with previous studies.\cite{sabine} $E_{total}$ seems to be only slightly higher for Germanium. Since the electronegativity of the metals is very similar, it is most likely attributed to the smaller covalent radius of germanium (Ge:122pm, Sn:140pm, Pb:146pm)\cite{GeSnPb} causing slightly shorter and stronger bonds. 

\begin{figure*}[htbp]
\centering
\noindent\includegraphics[width=0.96\textwidth]{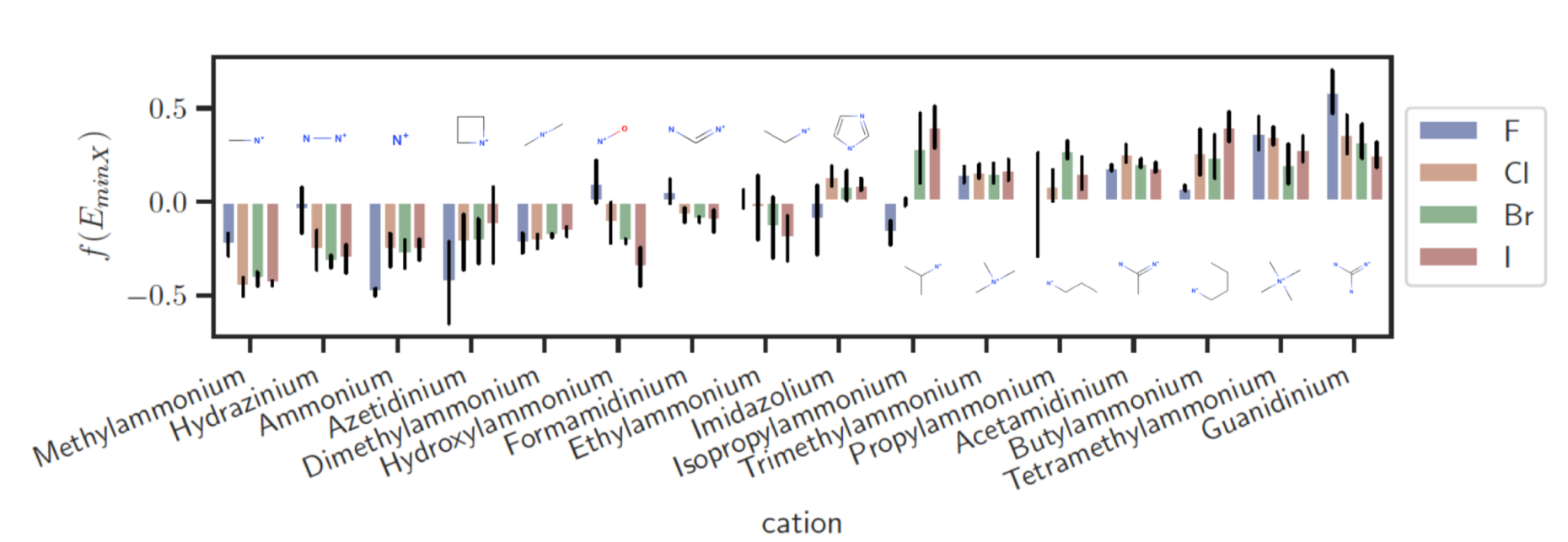}
\caption{A barplot showing the fractional deviations from the mean of the $E_{minX}$ values. The boxes show the standard deviation within the 3 points per metal. The cations are ordered by the mean of bars for Cl, Br and I. The cation structures are shown without their hydrogen atoms.}
\label{fpercation}
\end{figure*}

To compare the effect of the organic cation, we look at the fractional deviations of the individual $E_{minX}$ values with respect to the average of all the $E_{minX}$ values per halogen and metal. The fractional deviation  $f$ is defined as
$f(E_{minX})=\frac{E_{minX}-\overline{E}^{XM}_{minX}}{\overline{E}^{XM}_{minX}}$,
where $\overline{E}^{XM}_{minX}$ is the average of all the $E_{minX}$ values having halogen $X$ and metal $M$. The results are presented in Figure \ref{fpercation}.

It is remarkable that methylammonium, by far the most used in HOIPs, has the lowest $E_{minX}$ values, i.e., the weakest halogen-cation bond strengths, which well explains the degradation of MAPbI$_3$ to PbI$_2$ under UV light even without moisture or oxygen present.\cite{degr}
The observed trend can roughly be attributed to the size or the number of hydrogen atoms in the cation.
In Svane \latin{et al.}\cite{Svane2017}\ who studied 4 organic cations in formate perovskites, they found strong H-bonds in guanidinium and hydrazinium and weak ones in dimethylammonium and azetidinium. This agrees with the trend found here except for the hydrazinium cation, that has a remarkable low $E_{minX}$ value.

Svane \latin{et al.}\cite{Svane2017}\ also studied the HOIPs MAPbX$_3$ and FAPbX$_3$ with X=Cl, Br and I.
Remarkably, they found for FA slightly lower total electrostatic energies and hydrogen bonding energies than for MA.
We also find slightly lower hydrogen bonding energies for FA, but our results also further suggests that the $E_{total}$ values for FA are slightly higher than for MA because of $\pi\cdots\!X$ interactions (See supporting information section 1).
Interestingly, FA has the 3 but lowest $E_{total}$, while it is ranked the 6 but lowest for $E_{minX}$. This means that, despite FA has comparable total energy interaction as MA, it more evenly binds to the three different halogens, while MA leaves some halogens only weakly coupled.
The relatively higher ranking in $E_{minX}$ vs.\ $E_{total}$ is observed for all four cations having $\pi$-bonds.


\begin{table*}[htbp]
\small
\caption{Table giving the 5 structures with the lowest and the 5 with the highest cation-halogen bond strengths. Note that some structures appear in both columns: 
\textbf{7}=\textbf{18}, \textbf{8}=\textbf{19} and \textbf{10}=\textbf{20}. 
All values are in meV. The `M' and `X' columns give the type of metal and halogen respectively.
}
\label{tab:etotals}
\begin{tabular*}{\textwidth}{@{\extracolsep{\fill}}rcclr|@{\hskip .1in}|rcclr}
\hline
\#  & M & X & cation           & $E_{total}$ &
\#  & M & X & cation           & $E_{minX}$ \\ \cmidrule{1-5} \cmidrule{6-10}
\textbf{1}   & Sn  & I  & Hydrazinium      & 374  & \textbf{11}   & Pb  & I    & Methylammonium   &  92  \\
\textbf{2}   & Ge  & I  & Ammonium         & 380  & \textbf{12}   & Sn  & I    & Methylammonium   &  93  \\
\textbf{3}   & Pb  & I  & Hydrazinium      & 387  & \textbf{13}   & Pb  & I    & Hydroxylammonium &  95  \\
\textbf{4}   & Pb  & I  & Ammonium         & 396  & \textbf{14}   & Ge  & I    & Hydroxylammonium & 103  \\
\textbf{5}   & Sn  & I  & Ammonium         & 403  & \textbf{15}   & Ge  & I    & Methylammonium   & 104  \\ \cmidrule{1-5} \cmidrule{6-10}
\textbf{6}   & Ge  & F  & Hydroxylammonium & 2549 & \textbf{16}   & Ge  & F    & Propylammonium   & 769 \\
\textbf{7}   & Ge  & F  & Guanidinium      & 2712 & \textbf{17}   & Pb  & F & Tetramethylammonium & 796 \\
\textbf{8}   & Sn  & F  & Guanidinium      & 2777 & \textbf{18}   & Ge  & F    & Guanidinium      & 815 \\
\textbf{9}   & Pb  & F  & Acetamidinium    & 2804 & \textbf{19}   & Sn  & F    & Guanidinium      & 879 \\
\textbf{10}  & Pb  & F  & Guanidinium      & 2925 & \textbf{20}   & Pb  & F    & Guanidinium      & 931 \\ \hline
\end{tabular*}
\end{table*}

In Table \ref{tab:etotals} the 5 structure with the minimum and maximum $E_{total}$ and $E_{minX}$ are given.
Since the cation halogen bond strength decrease with decreasing electronegativity of the halogen is it logical that all minimum structures contain iodine while all maximum structures contain fluorine.
Guanidinium is the only cation that has both, a high $E_{total}$ and a high $E_{minX}$. 
Although Hydrazinium and Ammonium have the lowest $E_{total}$, Methylammonium and Hydroxylammonium have the weakest coupling to a halogen atom. 
Hydroxylammonium has the most noticable result since it binds very strongly to one halogen via an $O\!-\!H\!\cdots\!X\!-\!M$ bond, while it leaves other halogens only weakly bound.


\begin{figure}[h]
\begin{subfigure}{.23\textwidth}
  \centering
  \includegraphics[width=.7\linewidth]{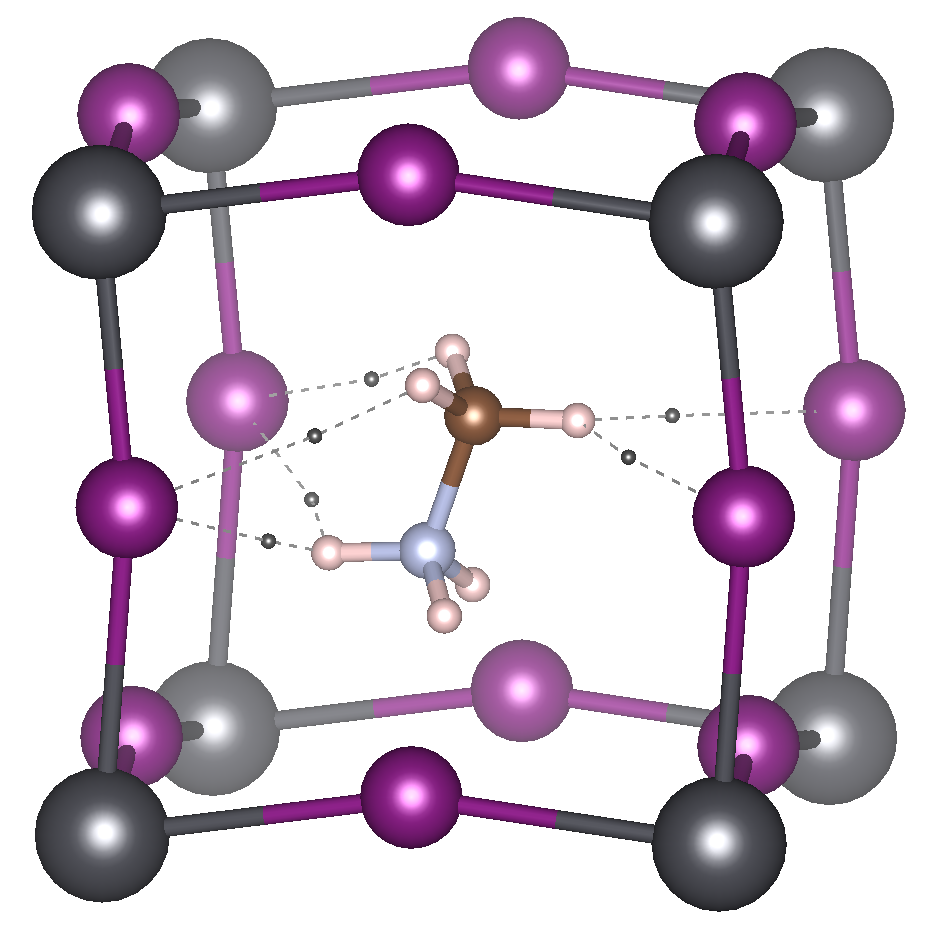}
  \caption{\textbf{11}}
\end{subfigure}%
\begin{subfigure}{.23\textwidth}
  \centering
  \includegraphics[width=.7\linewidth]{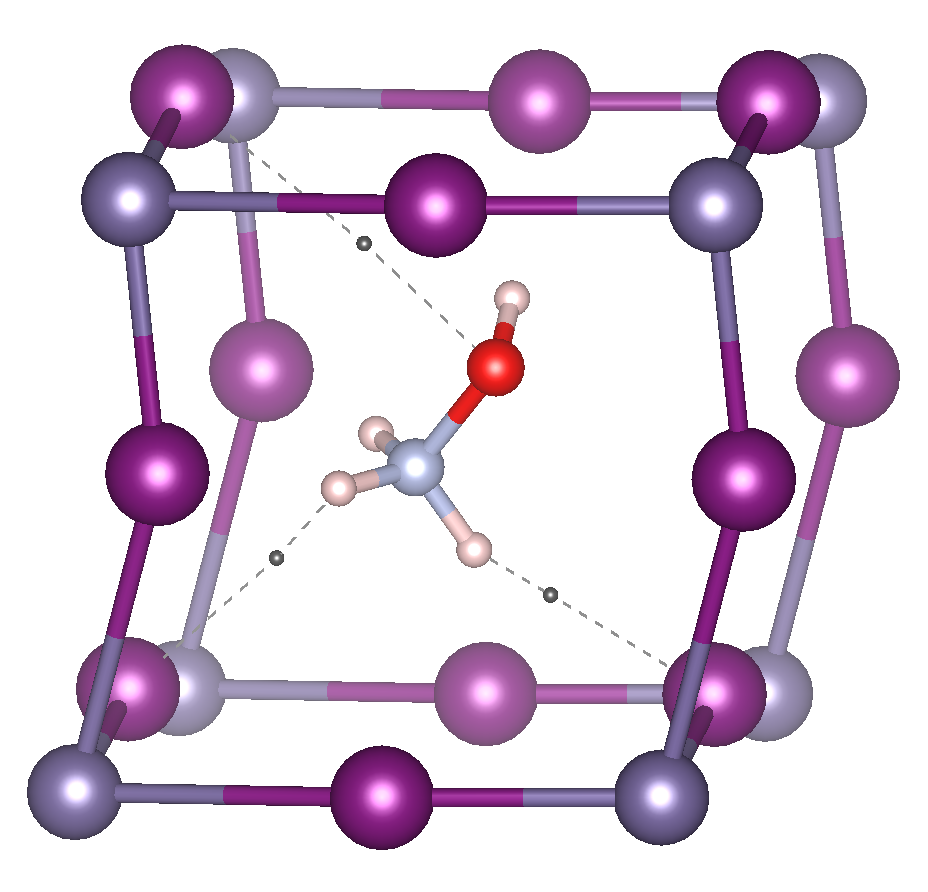}
  \caption{\textbf{14}}
\end{subfigure}
\begin{subfigure}{.23\textwidth}
  \centering
  \includegraphics[width=.7\linewidth]{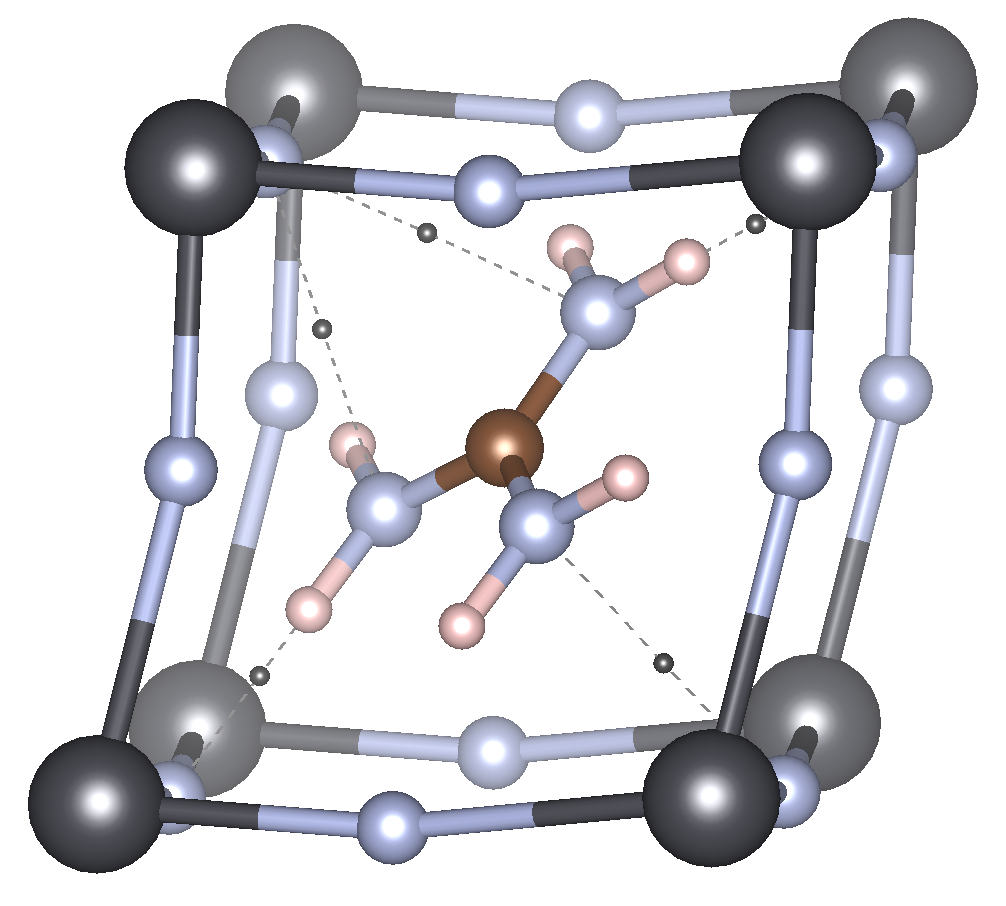}
  \caption{\textbf{20}}
\end{subfigure}
\begin{subfigure}{.23\textwidth}
  \centering
  \includegraphics[width=.7\linewidth]{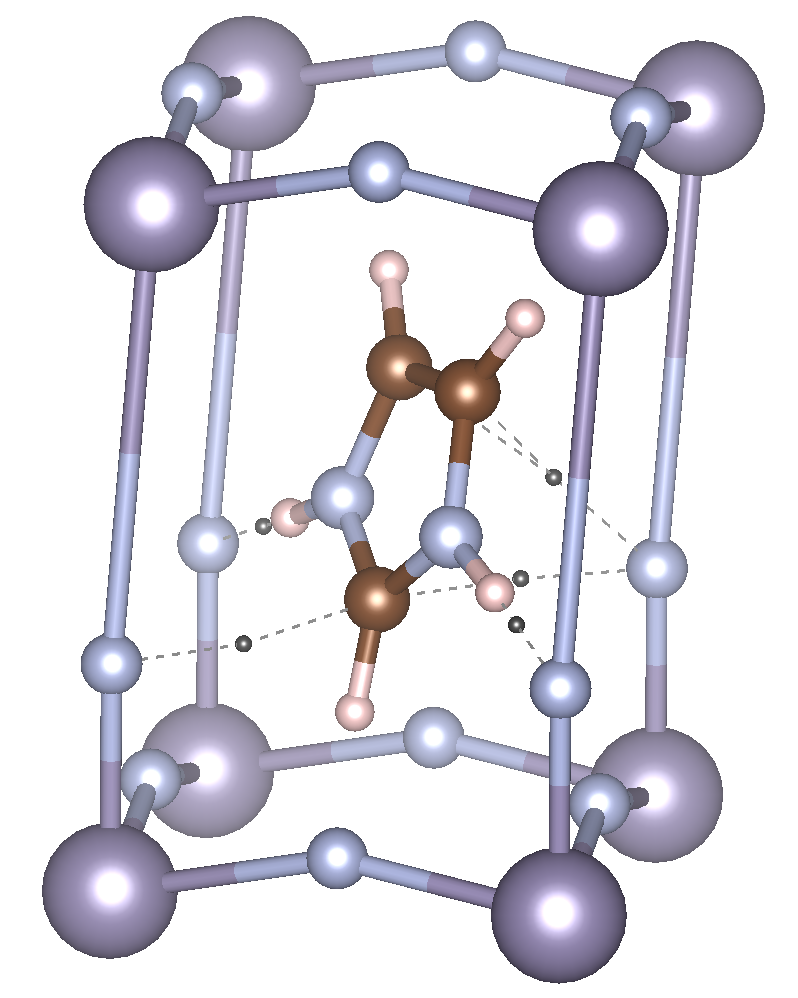}
  \caption{\textbf{21}}
\end{subfigure}
\caption{4 HOIP structures with the BCPs to only one type of halogen shown as little black dots, connected by dashed lines. The atoms are colored as Pb-black, I-purple, Ge-dark(steel) blue, F-blue, O-red, N-blue, C-brown and H-white.}
\label{fXminmax}
\end{figure}

Noticeable binding characteristics are found in specific structures. We discuss at first some structures with either very small or high $E_{minX}$. Next, we show the structures having the highest and lowest $E_{total}$ followed by a discussion of the structures that have the weakest and strongest individual bonds. 
By looking at these structures we can analyze the very nature of the bonds that contribute to $E_{minX}$ and what is the influence of the orientation of the cation.

In Figure \ref{fXminmax} only the BCPs that contribute to $E_{minX}$ are shown. 
Structure \textbf{11} (Methylammonium Lead Iodide), has the lowest $E_{minX}$ value of only 92 meV.
The weakest bound halogens are oriented approximately in the plane perpendicular to the C-N bond. Since neither the methyl, nor the amine group is oriented towards these halogens, the bond lengths are larger and the binding weaker. In total the $C\!-\!H\!\cdots\!I$ bonds account for more than 50\% of $E_{minX}$ illustrating the importance of these bonds in HOIPs (and MAPbI$_3$ in particular) as previously partially advocated by Varadwaj \latin{et al}.\cite{VARADWAJ2018, Varadwaj2019}
Another interesting structure with a very low $E_{minX}$ of only 103 meV is \textbf{14} (Hydroxylammonium Germanium Iodide).
The weakest bound halogens in \textbf{14} are only bound with two long hydrogen bonds of 2.9 and 3.4 \AA (together 86\% of $E_{minX}$) and a long chalcogen bond of 4.2 \AA($O\!\cdots\!F$).

Structure \textbf{20} (Guanidinium Lead Fluoride) has the highest $E_{minX}$ with two hydrogen bonds in the plane of the molecule that account together for 65\% of $E_minX$ and the other 35\% is caused by $\pi$-anion interactions thus adding a significant part of to $E_{minX}$.
Additionally, the structure of Imidazolium Tin Fluoride (\textbf{21}) is given in Figure \ref{fXminmax} because it has the halogens with the strongest interactions to the cation. The total binding energy to these halogen atoms in the whole dataset are 1290 meV, of which the two H-bonds account for 79\% and the rest are $\pi$-anion interactions.

\begin{figure}[h]
\begin{subfigure}{.23\textwidth}
  \centering
  \includegraphics[width=.8\linewidth]{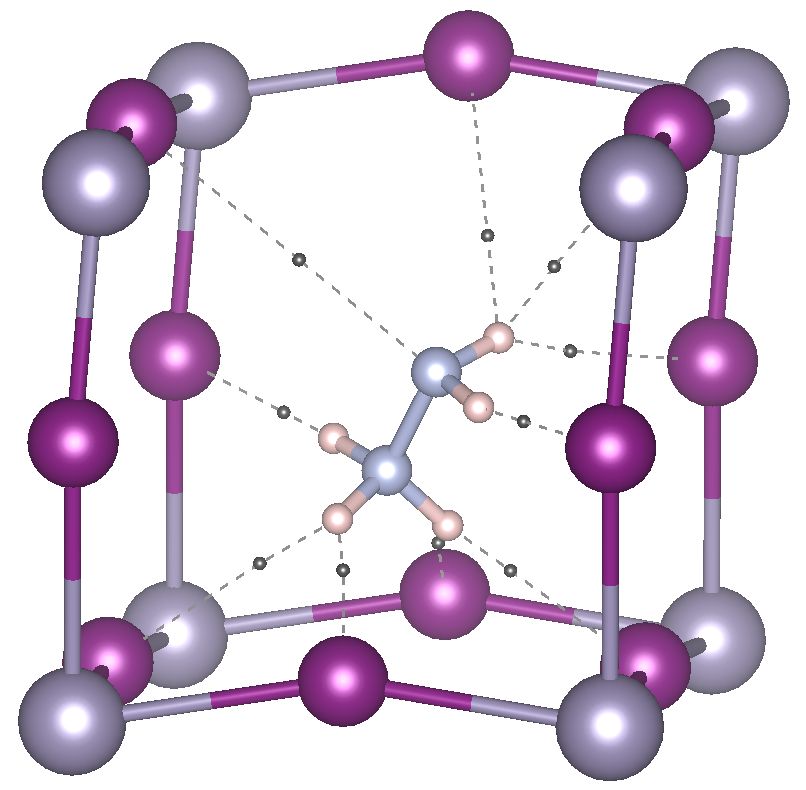}
  \caption{\textbf{1}}
\end{subfigure}%
\begin{subfigure}{.23\textwidth}
  \centering
  \includegraphics[width=.8\linewidth]{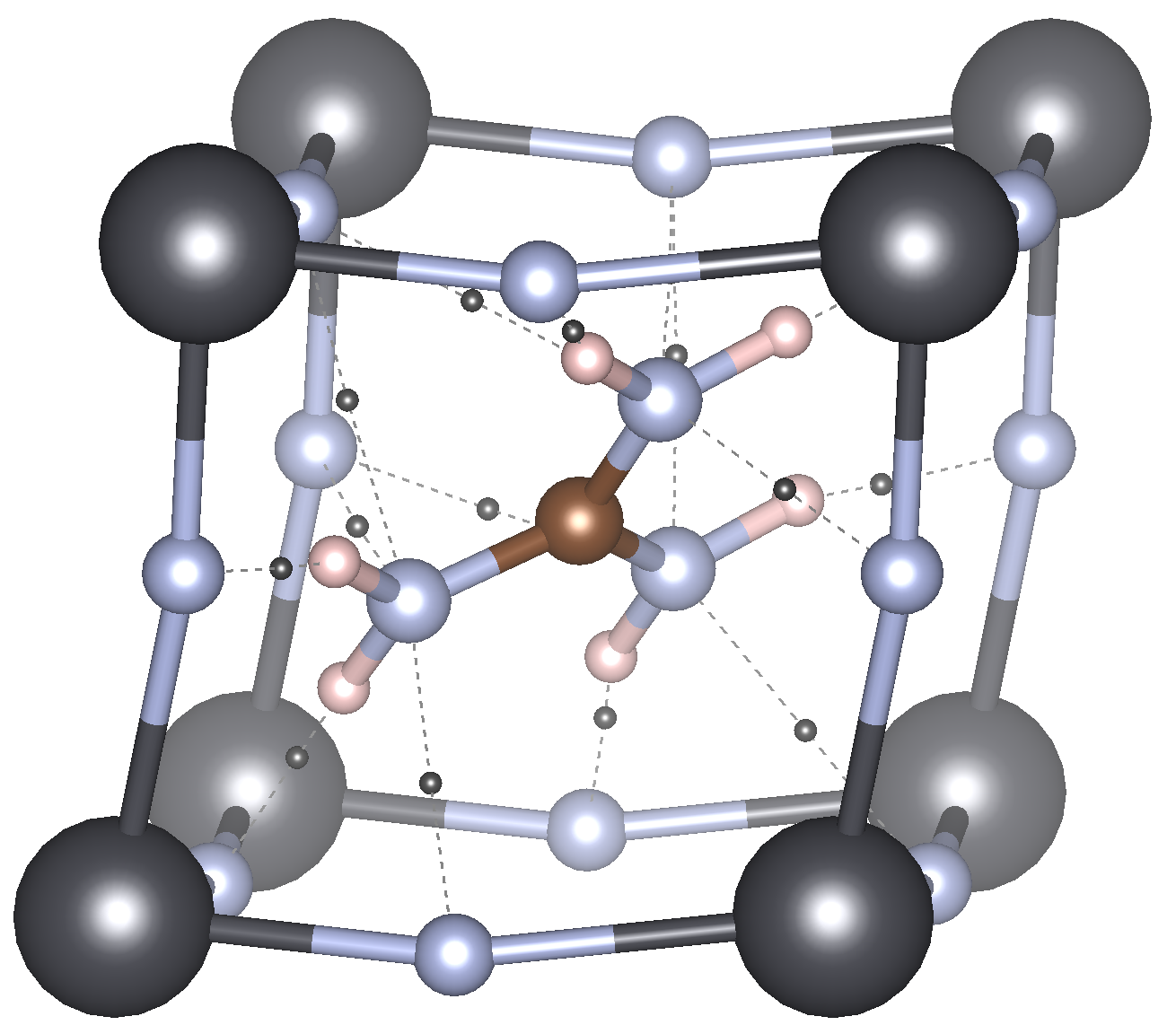}
  \caption{\textbf{10}}
\end{subfigure}
\begin{subfigure}{.23\textwidth}
  \centering
  \includegraphics[width=.8\linewidth]{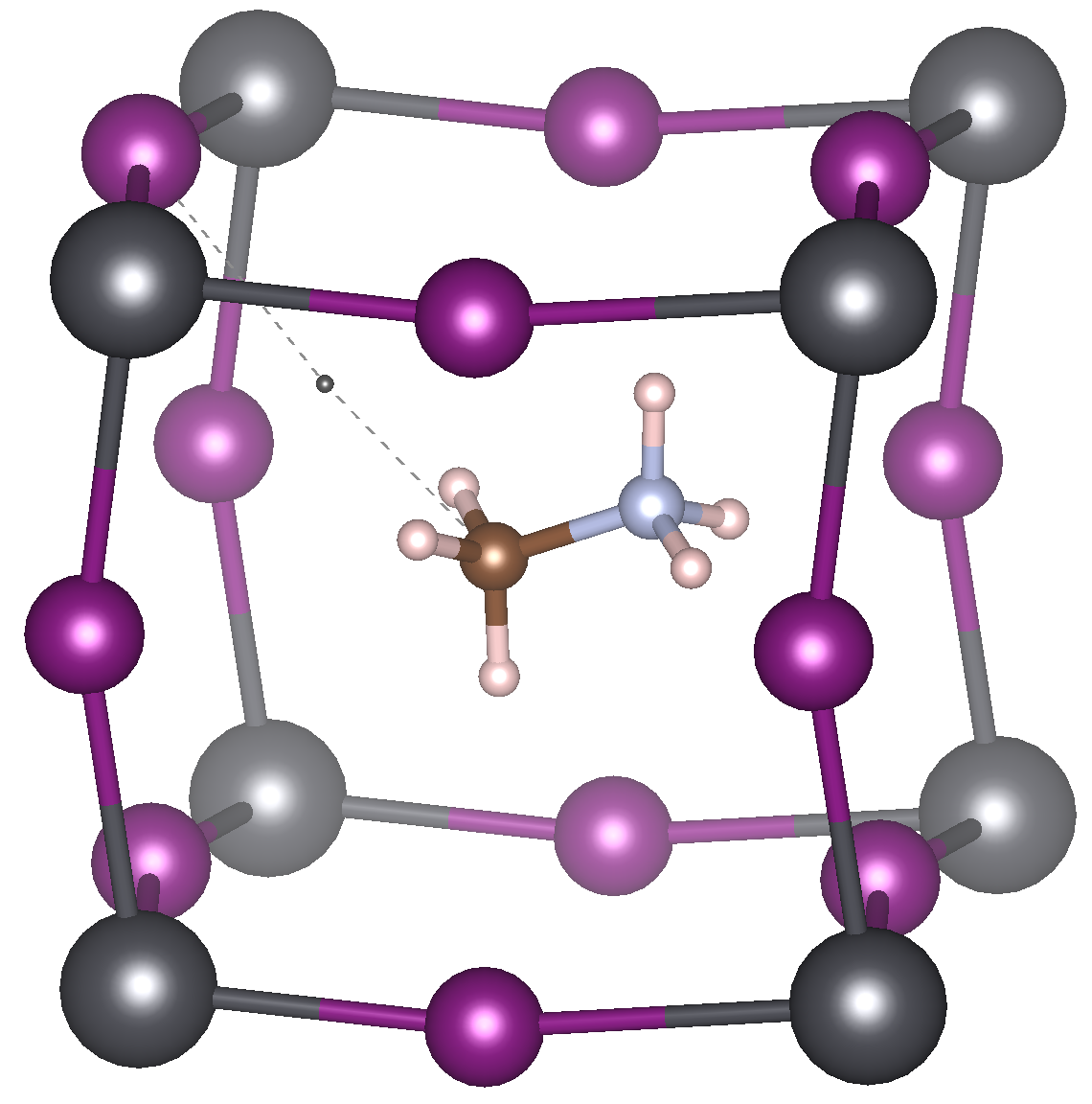}
  \caption{\textbf{11}}
\end{subfigure}%
\begin{subfigure}{.23\textwidth}
  \centering
  \includegraphics[width=.8\linewidth]{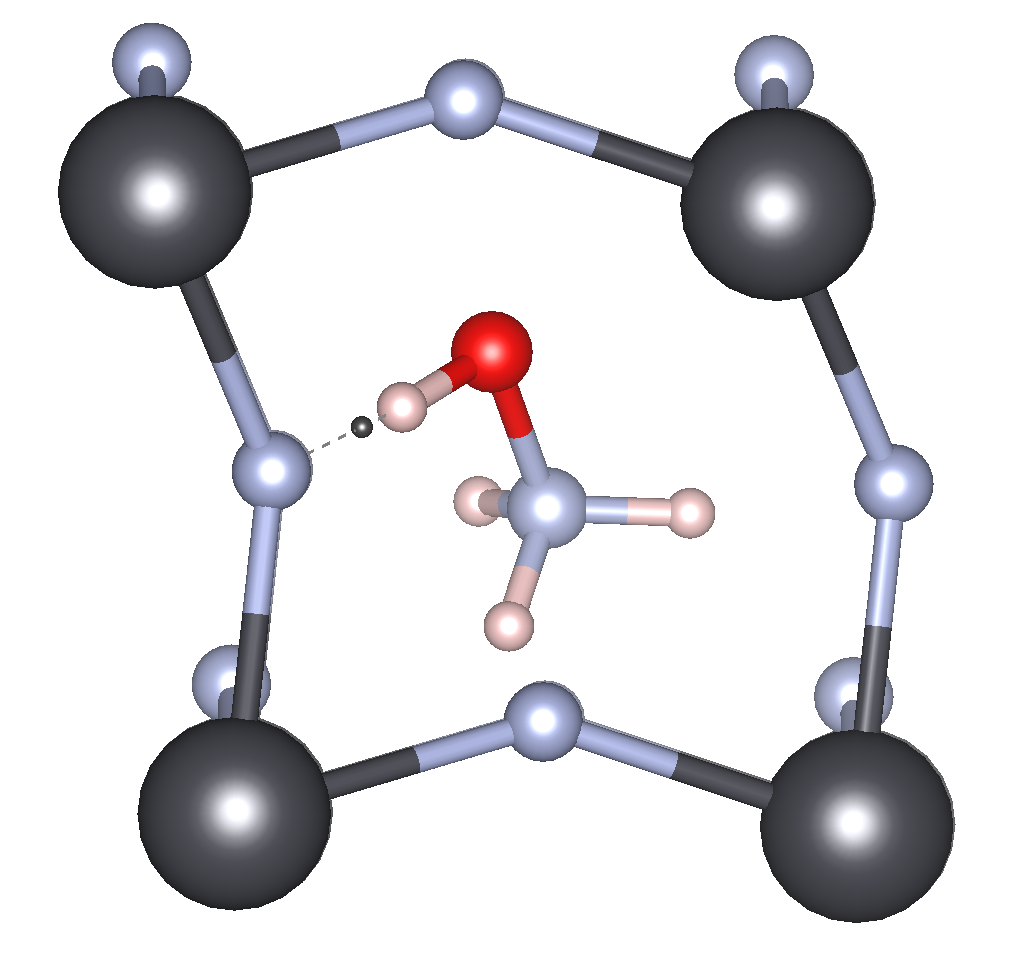}
  \caption{\textbf{22}}
\end{subfigure}
\caption{4 HOIP structures with the BCPs shown as little black dots, connected by dashed lines. The atoms are colored as Pb-black, I-purple, Ge-dark(steel) blue, F-blue, O-red, N-blue, C-brown and H-white.}
\label{fEtotalminmax}
\end{figure}

The structures with the lowest and highest $E_{total}$, \textbf{1} and \textbf{10}, (Hydrazinium Tin Iodide and Guanidinium Lead Fluoride), are shown in Figure \ref{fEtotalminmax}.
\textbf{1} has 10 BCPs while \textbf{10} has 15. It is surprising that $E_{total}$ for hydrazinium is even lower than for ammonium. This is most likely caused by the more favorable cation position in the ammonium structure where there are 3 strong non-bifurcated H-bonds while the fourth hydrogen of ammonium has a trifurcated H-bond, i.e., a single hydrogen participating in 3 H-bonds. This orientation is possible because ammonium can align its threefold rotational axis with the $<\!\!111\!\!>$ direction of the cube.
In structure \textbf{10}, 65\% of $E_{total}$ is caused by the strong H-bonds in the plane of the guanidinium molecule. Due to high symmetrically arrangement, guanidinium binds very evenly to all three types of halogens and thus $E_{minX} \lessapprox 3E_{total}$.
This provide a quantitative explanation of the exceptional binding properties of guanidinium in HOIPs. The combination of the strong H-bonds and the bonds not being free to rotate due to $\pi$-conjugation causes guanidinium to stay strongly bound at elevated temperatures as well \cite{Svane2017}.
Additionally note that guanidinium has a three-fold rotational axis aligned with the $<\!\!111\!\!>$ direction of the cube so it binds similarly with all three halogens.

In Figure \ref{fEtotalminmax}, also the structures with the weakest and strongest bond present in the dataset are depicted, as \textbf{11} and \textbf{22} respectively (Methylammonium Lead Iodide and Hydroxylammonium Lead Fluoride).
The weakest BCP is present in MAPbI$_3$ with only 7.7 meV.
This bond also corresponds with the BCP that has the lowest $\rho_0$ and $\nabla^2\rho$ values of 0.00128 and 0.0037 respectively.
This bond can be characterised as a carbon-halogen bond C$\cdots$I, a very weak so-called tetrel bond, only reported for MAPbBr$_3$ in one previous work\cite{VARADWAJ2018}.
The strongest bond in \textbf{22} is an O-H$\cdots$F bond.
From the whole dataset, this bond also is the shortest and has the highest density at its BCP (and second but highest density gradient). Note that this bond causes a large tilting angle in the perovskite. Large tilting angles often cause structural instability and non-photoactive phases.\cite{Wang2020, fedwa} 
Hence, to prevent large tilting angles, it is best for the cation to bind all the halogens strongly but evenly.

In summary, we can conclude that having a strong cation halogen bond does not mean having a high $E_{minX}$ as in hydroxylammonium, and that methylammonium has a low $E_{minX}$ due to the difficulty to couple to all three halogens while Guadinium binds very effectively.


\subsection{SISSO Results}

To increase our understanding of how the chemical composition, and in particular the cation structure influences the binding in HOIPs, we formulate here new mathematical descriptors composed from elementary topological features, that best correlate with $E_{minX}$.
To this end we apply the symbolic regression technique called SISSO, which designs mathematical descriptors by applying simple mathematical operators to the user-provided input values, i.e., the elementary "raw" features.
For the halogen and the metal the included elementary features are their polarizability $\alpha$, vdW radius $r$ and electronegativity $\chi$. 
The structure of the cation is encoded via its effective radius and relevant bond counts: the number of C-H, N-H, O-H and $\pi$ bonds. (A table of all features for all the structures can be found in the Electronic Supplementary Information.)
Note that to obtain this elementary features, no knowledge of the structures is required and evaluation of the descriptors should be as straightforward as evaluating the Goldschmidt tolerance factor.
The cation radii are only an approximation since most of the cations are not even close to being spherical. Nevertheless, they have been shown to be effective quantities within applications of predictive techniques on HOIPs.\cite{goldschmidt, ExtendTol, Kieslich2014}.

Since we already know that the type of halogen plays an important role, we ran SISSO in two ways: in the first case we apply SISSO on the whole dataset, named single-task. In the second case, we apply the multi-task learning capability of SISSO where the dataset is split into a subset for each halogen. 
Next, SISSO tries to find global descriptors while for each set different coefficients are optimized.
So, for a multi-set multi-dimensional SISSO run, $E_{minX}$ for a given HOIP $\mathcal{X}$ is:
$E_{minX}(\mathcal{X}) = c_i^0 + \sum_{j=1}^{N_{dim}} c_i^j d^j(\mathcal{X})$ 
where $c_i^0$ is an intercept of the $i^{th}$ task and $c_i^j$ is the coefficients of the $i^{th}$ task of the $j^{th}$ dimension and $d^j$ is the descriptor of the $j^{th}$ dimension, thus being identical for amongst all tasks.

\begin{figure}[htbp]
\centering
\noindent\includegraphics[width=0.5\textwidth]{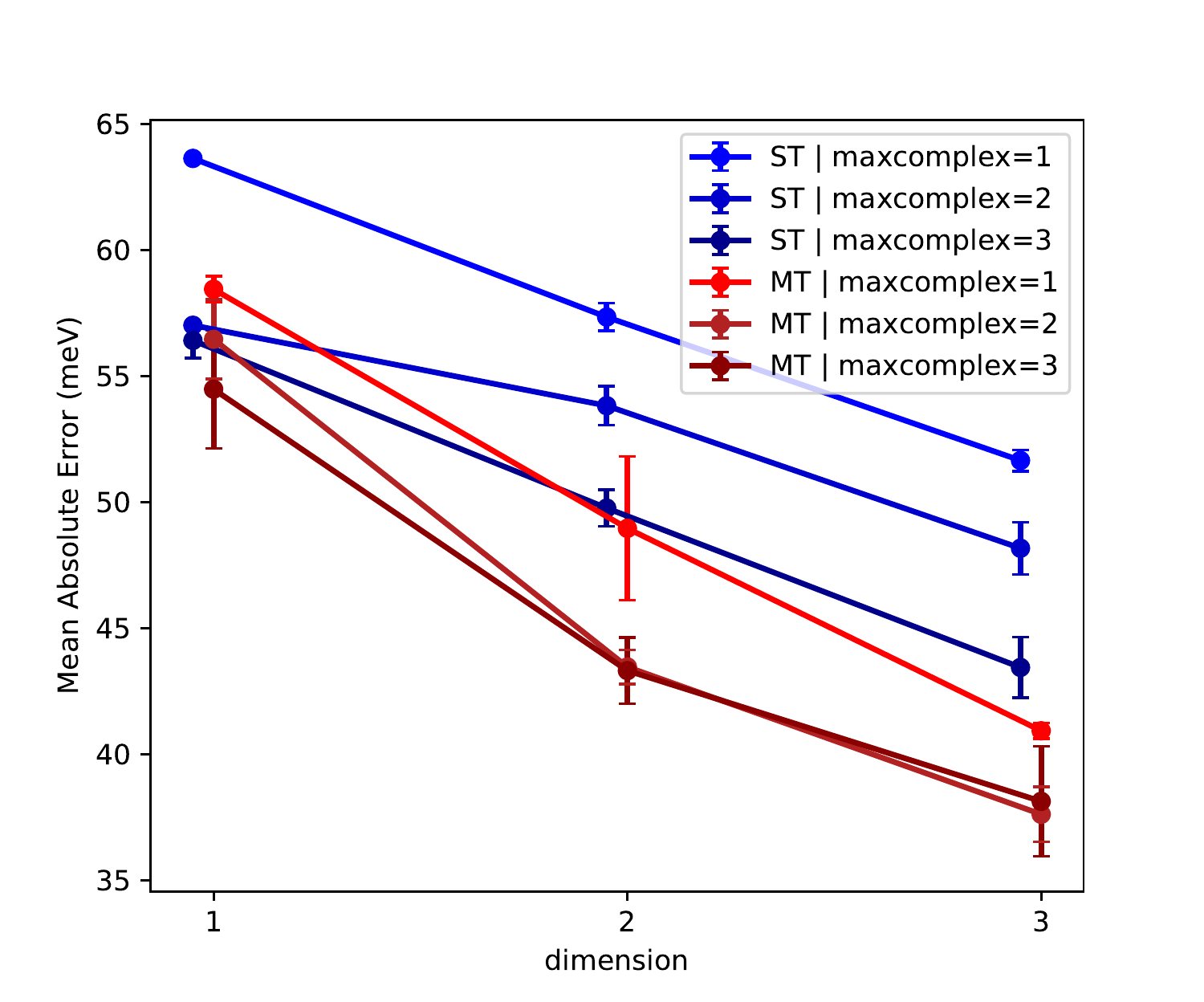}
\caption{SISSO Mean Absolute Errors vs dimension and maxcomplexity. ST=Single Task, MT=Multi Task.
For each setting, SISSO is run 5 times via cross-validation with test sets of 20\% of the data. The error bars indicate the minimum and maximum MAE of the 5 runs and the central dot is placed at the average MAE.
}
\label{fMAEs}
\end{figure}

To allow for scientific interpretability we apply a maximum to the descriptor complexity. This cutoff is effectively a hyperparameter to be chosen freely by the user. In Figure \ref{fMAEs} the errors as a function of dimension and complexity cutoff are shown for single and multi-task SISSO.
Multi-task SISSO always outperforms single-task SISSO, which is expected since it has to account for less variance within the subsets and there are 4 times more fitted coefficients.
Changing the complexity cutoff in Multi-task SISSO from 2 to 3 does not give any significant improvement.

{\renewcommand{\arraystretch}{1.5}%
\begin{table*}[htbp]
\small
\caption{Table showing the descriptors as found by SISSO setting the maximum complexity at 2. ST=single-task, MT=multi-task.
The descriptor coefficients are not shown but given in Supporting Information Section 3.
}
\label{tab:sisso}
\begin{tabularx}{\textwidth}{llYcYcY}
\cline{1-7}
dim                 & task & \multicolumn{5}{c}{descriptors} \\ \hline
\multirow{2}{*}{1D} & ST   &  $\frac{r_{cation}}{r_X^2}$   &       &     &      &     \\ \cmidrule{2-7} 
                    & MT   &  $N_{\pi} + \left| N_{CH} - N_{NH} \right|$                &       &     &      &     \\ \midrule \
\multirow{2}{*}{2D} & ST   &  $\frac{r_{cation}}{r_X^2}$    & ,     &  $\frac{N_{NH}*N_{\pi}}{\alpha_X}$   &      &     \\ \cmidrule{2-7}
                    & MT   &  $N_{CH}+N_{OH}+N_{\pi}$       & ,     &  $(N_{NH}+N_{OH})^2$   &      &     \\ \midrule
\multirow{2}{*}{3D} & ST   & $\frac{N_{OH}+N_{\pi}}{r_X}$   & ,     &   $\frac{N_{NH}+N_{OH}}{\alpha_X}$  & ,    &  $\frac{N_{CH}^2}{r_X^2}$   \\ \cmidrule{2-7} 
                    & MT   &  $N_{CH}+N_{OH}+N_{\pi}$       & ,     &  $N_{NH} + 2N_{OH}$                 & ,    &  
                    $N_{CH}*\left|N_{NH}-N_{\pi}\right|$   \\ \hline
\end{tabularx}
\end{table*}
}

The SISSO descriptors of single and multi-task are shown corresponding to the lines in Figure \ref{fMAEs} with maxcomplexity 2. 
In the single-task results, the halogen features are always present, accounting for the large difference in $E_{minX}$ for the different halogens. 
In the multi-task results the effect of the halogens is already taken into account via splitting the total dataset in 4 smaller sets, so the descriptors consider more the effect of the organic cation.
The results show that it is not only $N_{NH}$ that is important, but also the other bond counts $N_{CH}$, $N_{OH}$, and $N_{\pi}$.
The cation radius appears in the low dimension results, but disappears in the 3D result. This illustrates that the radius of the cation is not important \textit{per se}, but only inasmuch the radius of the cation correlates with the number of bonding partners of the cation.
The features accounting for the metal atom do not appear at all, indicating that the metal has no significant influence on $E_{minX}$.
It is remarkable that $N_{\pi}$ plays a dominant role in the SISSO descriptors for $E_{minX}$ as it is included in the multi-task 1D descriptor. 
These results further illustrate the importance of weak interactions for $E_{minX}$.

In the 3D multi-task descriptor a factor 2 appears. This can occur when SISSO uses the "$+$" operator twice with the same feature. Here, $N_{NH} + N_{OH} + N_{OH}$ simply reduces to $N_{NH} + 2N_{OH}$. 
It is a generally occurring phenomenon that SISSO descriptors are found that try to balance the importance of two elementary features by constructing "weights" and $N_{NH} + 2N_{OH}$ includes to elementary features $N_{NH}$ and $N_{OH}$ with weights 1 and 2 respectively.
In this case it is a logical consequence of the $O\!-\!H\!\cdots\!X$ bonds being stronger than the $N\!-\!H\!\cdots\!X$ bonds.

Still a significant part of the variance in $E_{minX}$ stays unexplained by the SISSO results, showing that other factors than those of the included topological features play a role. This is not a surprise, since the input features do not take into account the crystal packing. The packing factor is important since some cations can orient themselves more optimally than others to form strong non-covalent interactions to each halogen. As an example, the smallest cation, ammonium, can form a strong interaction with each of the tree types of halogens, explaining why its $E_{minX}$ is higher than those of methylammonium and hydrazinium.
Nevertheless, (given that the baseline MAE is 130 meV), it is remarkable that around 70\% of the variance can be explained using only these basic features, providing already great understanding in factors determining $E_{minX}$.


Last, we compare our results for the prediction of halides perovskite stability with the Goldschmidt Tolerance factor, commonly used to predict if a material will be stable in the perovskite conformation and given by:\cite{goldschmidt}
\begin{equation}
t = \frac{r_{cation}+r_X}{\sqrt{2}( r_M + r_X )}
\end{equation}
Although improved stability factors have been designed\cite{Bartel2019}, the general trend still applies that for a perovskite structure to be stable $t$ has to be approximately between 0.85 and 1.12 \cite{Burger2018, Li:wf5033}.
Our results show that roughly $E_{minX}$ and $E_{total}$ increase with larger cations, while the Goldschmidt factor puts a limit at how large these cations can be and a balanced solution has to be chosen.
So, although our results show that HOIPs will become more structurally stable w.r.t.\ halogen migration when the cation binds more strongly to the halogens, the perovskite structure itself might become unstable when larger cations are used. For example, Guanidinium Germanium Iodide has $t=1.2$, thus lying outside the stability window, while Guanidinium Lead Iodide, having a larger metal, is stable ($t=1.03$).
Nevertheless, also small additions of other cations might improve the perovskite stability while keeping the Goldschmidt factor within the acceptable range.
When $t$ is too high, the tilting angles of the MX$_6$ octahedrons increase leading to non-photoactive phases.
For example, at room temperature FAPbI$_3$ exist in a non-photoactive phase while when FA is mixed with MA or Cs, the photoactive phase becomes stable.\cite{Burger2018, Ava2019}


For what concerns the most common cations (MA and FA), our results generalize a previous study \cite{oranskaia2019} where only the MA and FA cations were compared, showing higher binding strengths for the FA cation.
Indeed, our findings generalize the higher $E_{total}$ of FA than MA to any type of halogen or metal, but by looking at $E_{minX}$ instead of only looking at $E_{total}$, our work makes an even stronger case to use FA instead of MA, since MA binds some halogens particularly weakly.


\begin{figure}[htbp]
\centering
\noindent\includegraphics[width=.5\textwidth]{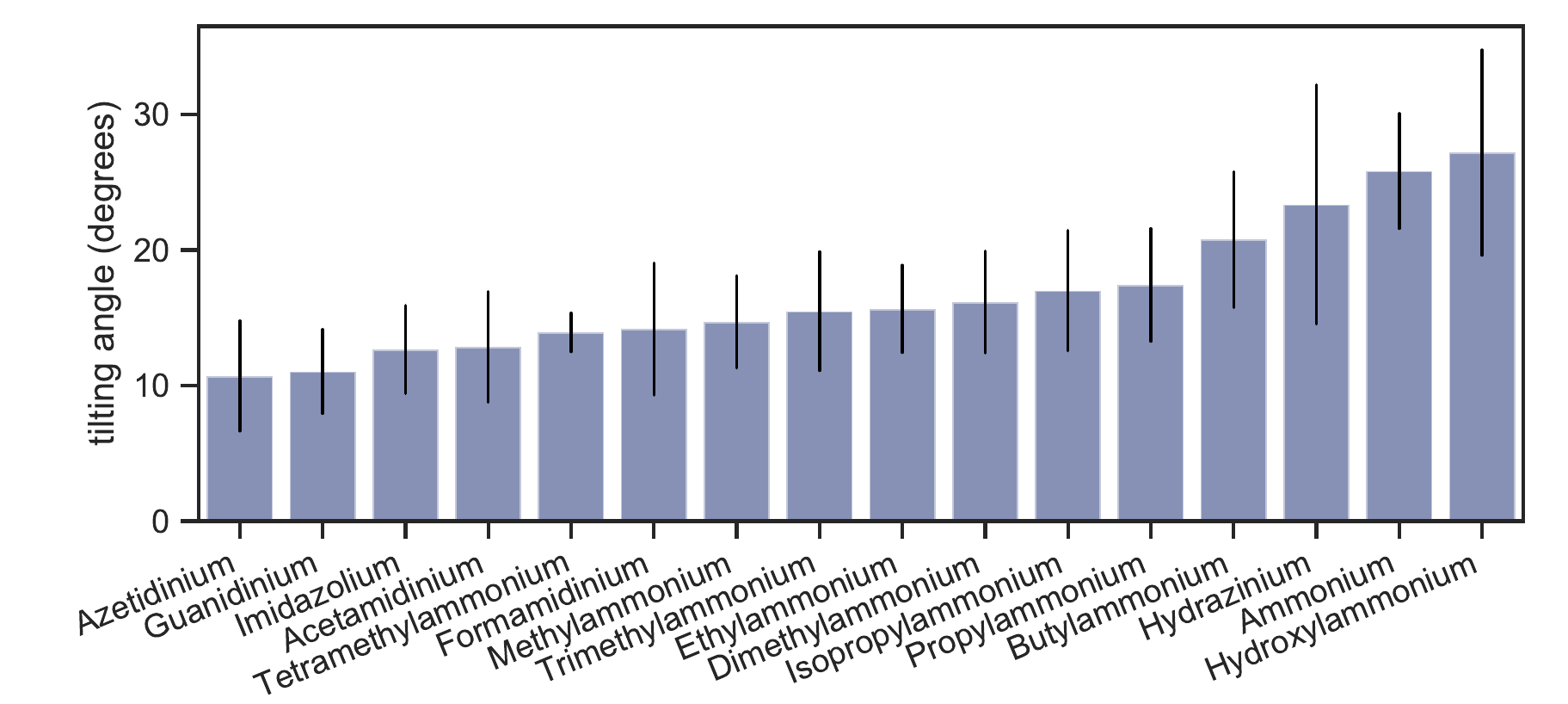}
\caption{Average tilting angles per cation}
\label{ftilts}
\end{figure}


In Figure \ref{ftilts} the average tilting angles are given for each organic cation.
The extent of octahedral tilting is \textit{a complex phenomenon that is driven by the interplay between many interactions}.\cite{Varadwaj2019}
Bernasconi \latin{et al.}\cite{Bernasconi2018}\ found that the tilting angle in MAPbX$_3$ (with X is Cl, Br and I) correlates with the strongest hydrogen bond. 
As one can see from structure \textbf{11} in MAPbI$_3$ the tilting is oriented towards the ammonium group. 
This trend is also more generally observed as the tilting angle is the largest for Hydroxylammonium since it forms a very strong hydrogen bond (see structure \textbf{23}). 
Yet, the trend is more intricate since the cation with the second highest tilting is ammonium.
The large tilting observed for ammonium can be attributed to the small size of the cation and also to the strong hydrogen bonding of ammonium to three halogens, pulling these halogens closer. 
Additionally large tilting angles are observed for structures with an ammonium group and an additional alkane group (methyl-, ethyl-, propyl-, isopropyl-, tetrabutyl- and dimethylammonium), all structures where the stronger hydrogen bonds of N-H groups on one side of the cavity creates a pronounced tilting effect.
The smallest tilting angles are observed for HOIPs with the Azetinidinium cation. Here the two N-H groups only bind to one type of halogen, while the other two types of halogens are binding exactly evenly by all C-H groups, thus giving an almost zero tilting angle for two dimensions and only a slight tilting in the dimension where the two N-H groups bind with the halogens.
Small tilting angles are also observed for Guanidinium since it binds very evenly to all the 12 halogens of the cavity.


Our results show that their is a meaningful relation between the bond counts of the cation and the HOIP stability. 
While former Machine Learning models are only based on effective radii and electronegativity values,\cite{WU2019, ExtendTol, Pilania2016}
from our results it can be induced that inclusion of the bond counts, especially when combined like in the SISSO descriptors, express useful quantities.
For organic cations, these bond counts give far more information than only the effective radii since the presence of $\pi$ bonds or O-H bonds is crucial information not captured in only an effective radius.
Specific bond counts affect the non-covalent interactions, which dictate tilting angles, that then have an influence on carrier mobility and band gaps.

\section{\label{sec:4}Conclusions}


HOIPs represent a potential game-changer technology to harvest solar energy. Still, the intrinsic instability and degradation due to external factors of such systems prevent large scale production. 
The present study has identified the key aspects that contribute to the stabilization of halide ions in HOIPs, the major species that undergoes migration in absence and under the effect of different external factors, creating instability and degradation.

Our work has revealed the fundamental bond and composition properties that most affect the halogen-cation interactions in hybrid perovskites in a large chemical space. This allows us to give a rationale of previous results, based on smaller spaces and to highlight previous misconceptions and overlooked aspects.
In particular, we unveil the reasons of an improved stability given by the halogen F, the origin of the higher stability offered by FA compared to MA, which shows to be different from previous explanations in the literature, and the reason behind the exceptional case of overall stronger bonds for guanidinium. A clear inverse power law between bond lengths and density at the BCPs is also found. Our results highlight the importance of the cation-halide interactions and the urge to differentiate between different halide positions. We have corroborated in a quantitative and first-principles manner the importance of non-covalent interactions other than $N\!-\!H\!\cdots\!I$, such as $C\!-\!H\! \cdots\!I$ and $N/C\!\cdots\!I$ for MAPbI$_3$, advocated in a previous work\cite{Varadwaj2019}, confirming that the neglect of such interactions in the majority of studies may lead to potentially wrong criteria for the design of new HOIPs. We did this not only for MA, but for a whole array of 16 different organic cations. 
Additionally in some cations such as FA, also $\pi\!\cdots\!X$ are found. 
We show that these weak interactions have a significant role to bind the halogen atoms, and thus are expected to be important to prevent halogen migration. Therefore, the in-depth understanding of weak non-covalent interactions is fundamental in the molecular design of new more stable HOIPs.
Moreover, as the most prevalent cation MA even has the lowest $E_{minX}$, there is ample room for cation substitution to increase the cation halogen coupling.

We found new mathematical, physically meaningful descriptors that identify the factors that dominate the coupling between the halogens and cations. We showed that the factors in order of importance are the type of halogen, the presence of $\pi\!\cdots\!X$ bonding and the number of hydrogen bond donors including $C\!-\!H$. Bonds other than H-bonds have been overlooked in previous studies, like the considerable importance of $\pi$-bonds in binding to the weakest halogen bonds. We show that bond counts are more meaningful than only the effective radii as used in the prevalent Goldschmidt tolerance factor. Moreover, the found descriptors unveil the importance of $\pi$-interactions for strong cation-halide interactions.
Based on this work, we suggest HOIP stabilization by inclusion of larger cations, preferably those having $\pi$-bonds. Guanidinium proves to be the most effective cation due to multiple $N\!-\!H$ hydrogen bond donors, a large $\pi$-conjugated system and its efficient packing due to alignment with the $<\!111\!>$ direction, thus binding evenly with all halogens.

Our work focuses on enhanced stability via a robust design approach for strengthening molecular bonds, and offers an alternative route to the different approaches previously used for HOIPs. We believe that our work will motivate the investigation of new hybrid perovskite materials with strong and balanced non-covalent interactions. The new descriptors can be used to train machine learning algorithms for even larger chemical spaces, to drive the design of HOIPs with chemical bonding enhancement and immobilization of the potentially migrating ions. As our study focuses on shaping atomic-scale properties within one unit cell, it potentially offers the possibility to design large size single crystals that do not segregate and do not form grains.

\begin{acknowledgement}

This work has received funding from the Research Executive Agency under the EU's Horizon 2020 Research and Innovation program ESC2RAD (grant ID 776410). 
We are grateful to PRACE for granting us access to the computational resources of the Joliot-Curie SKL at CEA (France) (PRACE Project: 2019215186)
and we thank the Space POLE HPC for providing the resources for the SISSO calculations.

\end{acknowledgement}

\begin{suppinfo}

The Supporting Information document shows a plot of fractional mean deviations from $E_{total}$ instead of $E_{minX}$. Also a boxplot and barplot for the total set of 548 perovkite structured HOIPs in the dataset is presented. Additionally the SISSO intercepts and coefficients are given. Finally a section about the correlations between bond lengths and the bond critical points is presented.

Two additional files are available. The first one is a comma-separated file listing all the 548 HOIPs adhering to a perovskite structure with their continuous symmetry measures for cubicity, $E_{total}$, $E_{minX}$ and $E_{atomozation}$ values. The second file is a SISSO input file listing all the input features.

See DOI: 00.0000/00000000.

\end{suppinfo}

\bibliography{citations}

\newpage
\section{TOC Graphic}
\includegraphics[width=.5\textwidth]{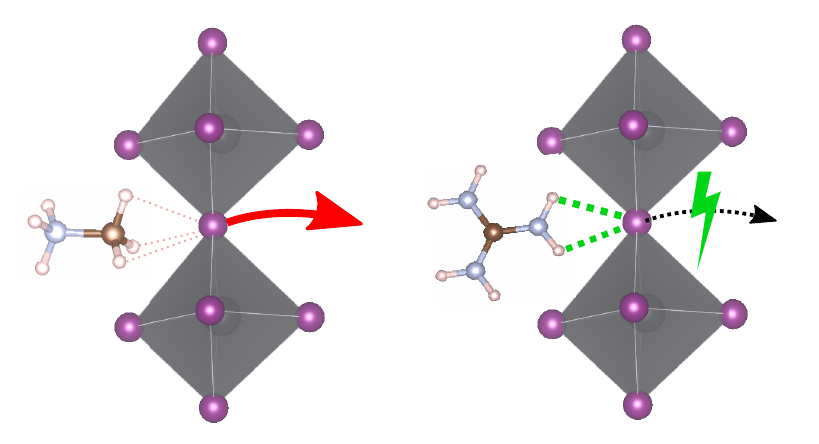}

\end{document}


\title[]{Supporting Information for Publication for "Molecular Bond Engineering and Feature Learning for the Design of Hybrid Organic-Inorganic Perovskites Solar Cells with Strong Non-Covalent Halogen-Cation Interactions"}

\author{Johannes L. Teunissen}
\email{johannes.teunissen@aeronomie.be}
\affiliation{Royal Belgian Institute for Space Aeronomy, Av Circulaire 3, 1180 Brussels,
Belgium}
\affiliation{Univ. Lille, CNRS UMR 8520 - IEMN - Institute of Electronics, Microelectronics and Nanotechnology, F-59000 Lille, France}

\author{Fabiana Da Pieve}
\affiliation{Royal Belgian Institute for Space Aeronomy, Av Circulaire 3, 1180 Brussels,
Belgium}
\date{\today}

\maketitle

\section{Plot of fractional mean deviations from $E_{total}$ instead of $E_{minX}$}

\begin{figure}[H]
\centering
\noindent\includegraphics[width=\linewidth]{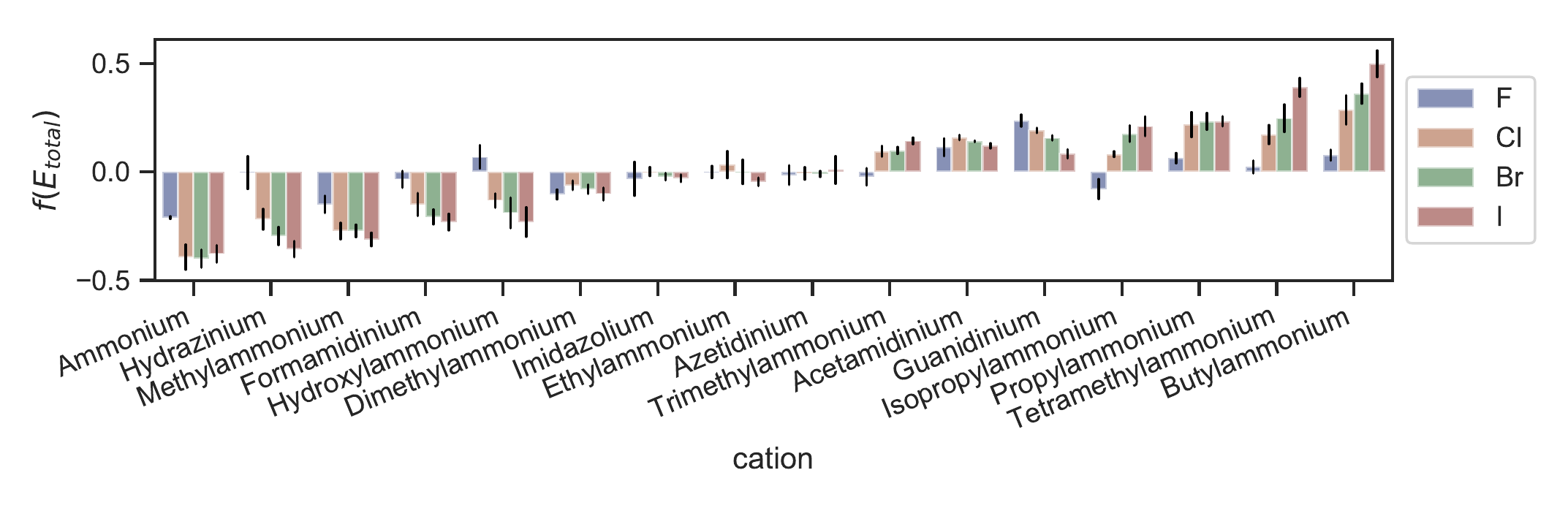}
\caption{A barplot showing the fractional deviations from the mean of the $E_{total}$ values. The boxes show the standard deviation within the 3 points per metal. The cations are ordered by the mean of bars for Cl, Br and I}
\label{fpercationEtotal}
\end{figure}



\newpage
\section{Boxplot and barplot for total set of perovkite structured HOIPs}

\begin{figure}[H]
\centering
\noindent\includegraphics[width=12cm]{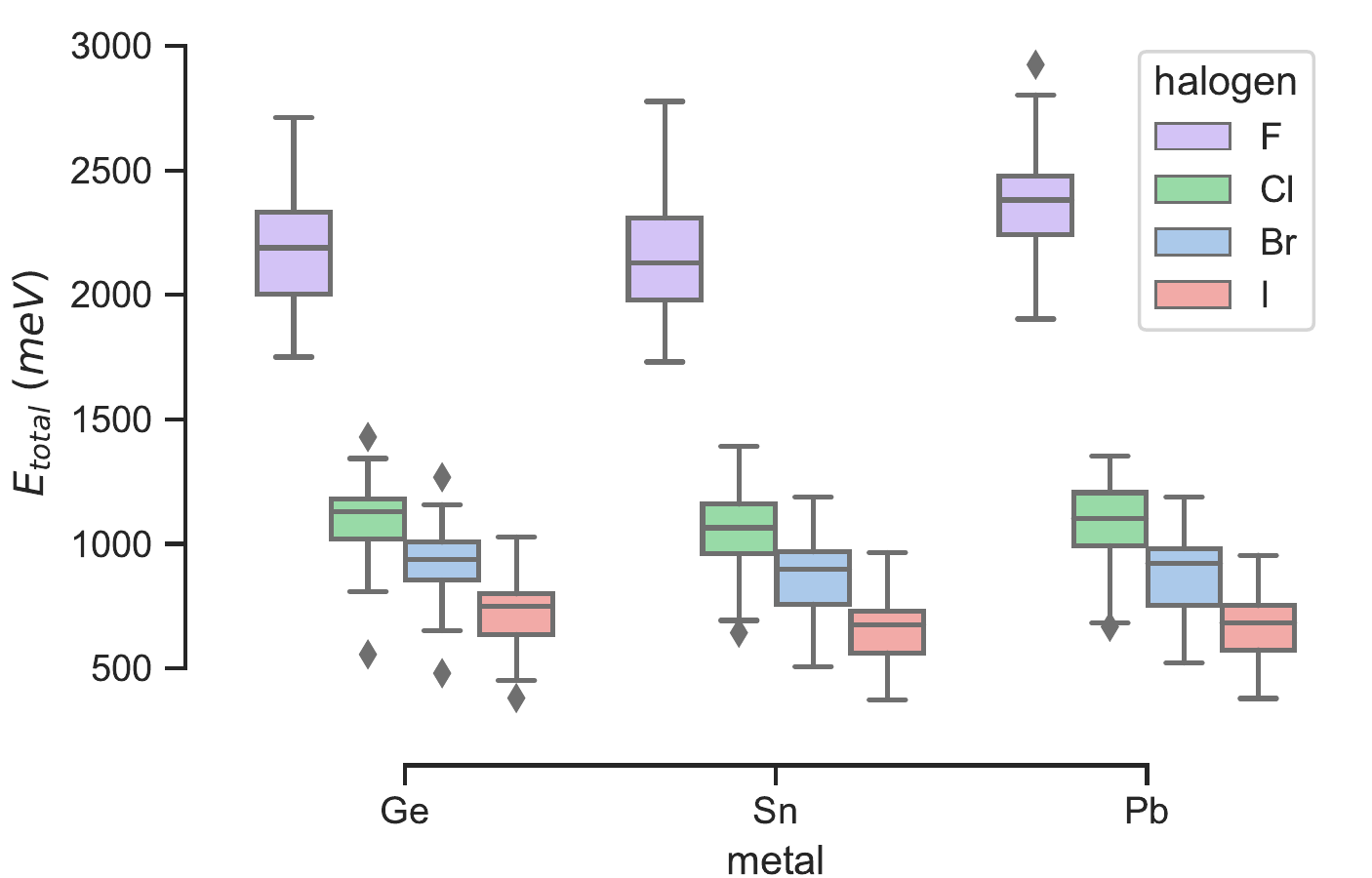}
\caption{A boxplot showing the distribution of the $E_{total}$ values grouped by metal and halogen. The boxes show the quartiles of the dataset while the whiskers are extended to 1.5 times the interquartile range (IQR).}
\label{fboxplot548}
\end{figure}

\begin{figure}[H]
\centering
\noindent\includegraphics[width=\linewidth]{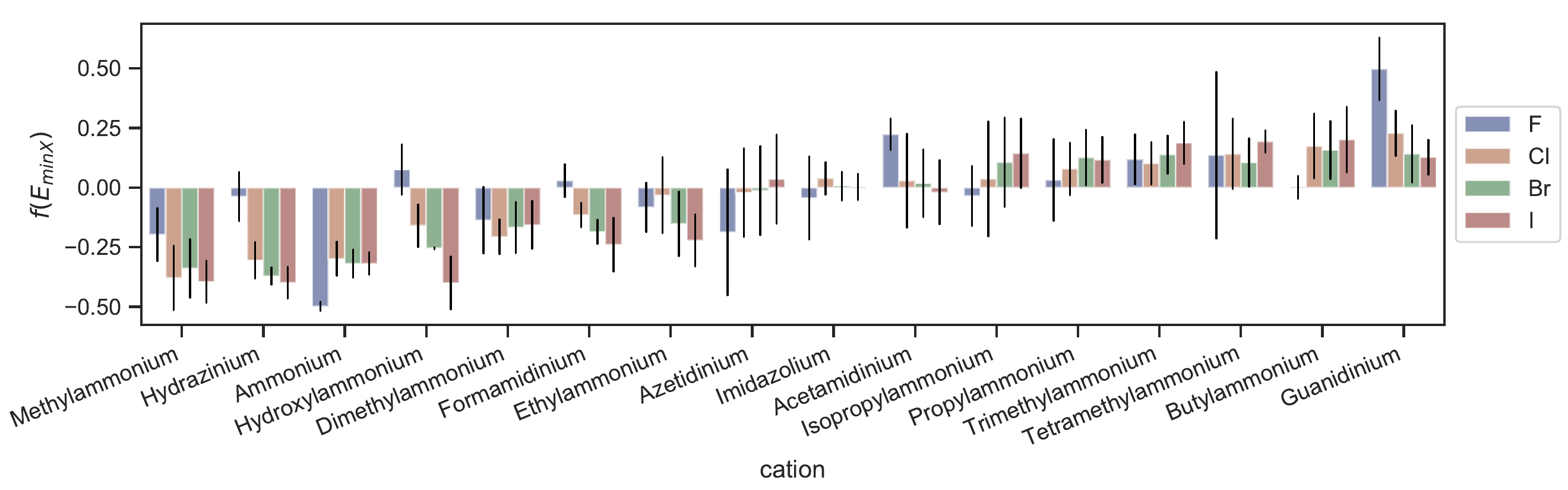}
\caption{A barplot showing the fractional deviations from the mean of the $E_{minX}$ values. The boxes show the standard deviation within the 3 points per metal. The cations are ordered by the mean of bars for Cl, Br and I}
\label{fpercation548}
\end{figure}




\newpage
\section{SISSO incepts and coefficients}

When the units are right, the product of the coefficients with the descriptor should yield the same unit as the target property, here an energy value in meV. In order to obtain the right units, implicitly every coefficients has the same unit as the target property multiplied by the inverse of the descriptor unit. 
For example, for the descriptor $\frac{r_{cation}}{r_X^2}$ the unit is $\angstrom^{-1}$ and the unit of the corresponding coefficient will be $meV \cdot (\angstrom^{-1})^{-1} = meV\angstrom$

\begin{table}[H]
\caption{}
\label{tab:my-table}
\begin{tabularx}{\textwidth}{cccYYYY}
\hline
 &  &  &  & \multicolumn{3}{c}{descriptor coefficients} \\ \cline{5-7} 
dimension & type & task & intercept & 1st & 2nd & 3rd \\ \hline
\multirow{5}{*}{1D} & ST &  & -10.83775855 & 319.9727192 &  &  \\ \cline{2-7} 
 & \multirow{4}{*}{MT} & F & 388.8449501 & 31.40147593 &  &  \\ \cline{3-7} 
 &  & Cl & 201.731012 & 15.29598431 &  &  \\ \cline{3-7} 
 &  & Br & 169.2543037 & 12.25954921 &  &  \\ \cline{3-7} 
 &  & I & 123.5074933 & 9.572764922 &  &  \\ \hline
\multirow{5}{*}{2D} & ST &  & -15.32977489 & 320.2275489 & 44.64510666 &  \\ \cline{2-7} 
 & \multirow{4}{*}{MT} & F & 133.2169537 & 45.49644431 & 16.48683707 &  \\ \cline{3-7} 
 &  & Cl & 72.46005121 & 24.79553435 & 6.465460895 &  \\ \cline{3-7} 
 &  & Br & 59.21246324 & 21.10295285 & 5.30557354 &  \\ \cline{3-7} 
 &  & I & 32.32262889 & 17.45283407 & 4.031622733 &  \\ \hline
\multirow{5}{*}{3D} & ST &  & 85.81368442 & 58.70445839 & 355.0621774 & 5.719060367 \\ \cline{2-7}
 & \multirow{4}{*}{MT} & F & −152.4998827 & 76.19853998 & 131.4068389 & -12.37686022 \\ \cline{3-7} 
 &  & Cl & −42.50024164 & 34.82282038 & 52.5412850 & -3.190133102 \\ \cline{3-7} 
 &  & Br & −50.71065481 & 28.48207304 & 44.04192996 & -1.073078228 \\ \cline{3-7} 
 &  & I & −47.58416835 & 22.38645474 & 32.65747739 & -0.0311616557 \\ \hline
\end{tabularx}
\end{table}

\newpage
\section{Correlations between bond lengths and bond critical points}

\begin{figure}[htbp]
\centering
\noindent\includegraphics[width=0.7\textwidth]{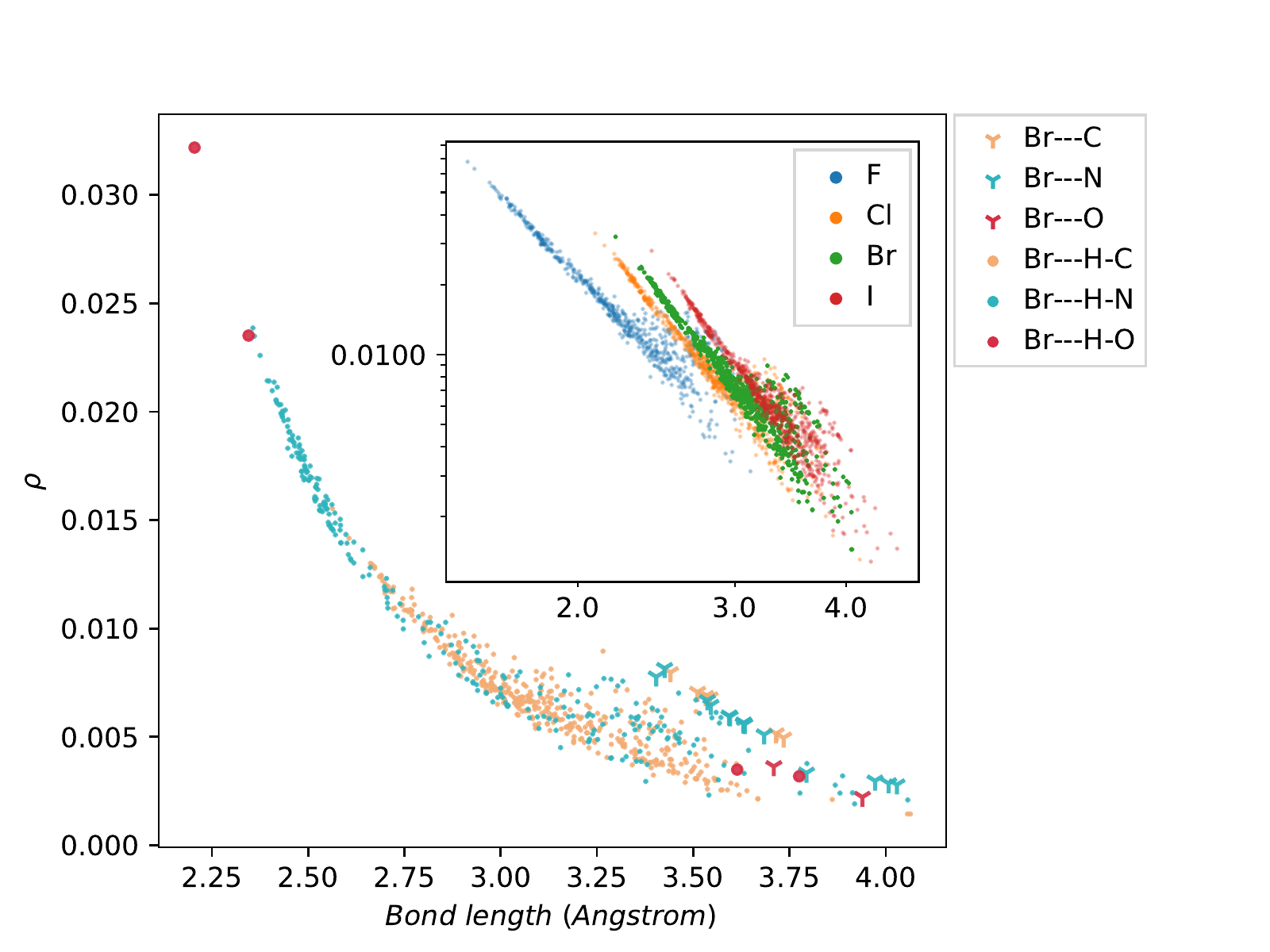}
\caption{Two scatter plots of the densities at the BCPs vs the bond lengths. The inset shows all values, colored by the type of halogen. The main plot shows only the values corresponding to Bromine, colored by the bond type. (Hence the green points in the inset are the same points as all the points in the main plot.) The axes of the inset belong to the same properties as the main plot, but the axis scales are logarithmic.}
\label{fBLvsE}
\end{figure}

In Figure \ref{fBLvsE} the densities at the BCPs are plotted vs.\ the corresponding bond lengths. The main plot only reports the results for the chosen case of bromine, for sake of clarity, as the breaking down of this data per halogen to the different type of bonds (shown in the inset) follows a very similar patterns for each halogen. 

For the main plot, from left to right, there are first the short and high density Br$\cdots$H-O bonds followed by the onset of the 
Br$\cdots$H-N and Br$\cdots$H-C bonds following the expected trend that weaker bonds are longer. 
The two Br$\cdots$H-O bonds found at bond lengths higher than 3.5 \AA are bifurcations of the strong bonds at small bond lengths.
This is a pattern that is more generally found in HOIPs were hydrogen atoms form one strong H-bond and also participate in some much weaker bonds.\cite{Svane2017, VARADWAJ2018}

Although the majority of the bonds are H-bonds, there are 6 cations that form bonds without hydrogen, namely the previously mentioned four cations with $\pi$ bonds and hydrazinium and hydroxylammonium; the latter two have a lone pair of nitrogen and two lone pairs of oxygen respectively. All these bonds are shown in Y-shaped symbols in Figure \ref{fBLvsE}. These bonds can be seen as chalcogen, pnictogen or $\pi$-anion binding.\cite{anionpi}

As can be seen in the inset plot of Figure \ref{fBLvsE}, only some bonds with fluorine have densities at the BCP that transcend outside the density range for weak interactions, thus illustrating their strong bonding. 
Generally, the H-bonds in HOIPs have a strong electrostatic(ionic) component due to the strong $\delta^-$ on Fluorine and the negative charge of the cation\cite{VARADWAJ2018}, making these exemplary for charge-assisted H-bonds (CAHB) of the strongest type with the H-bond acceptor negatively charged and the H-bond donor positively charged, i.e., CAHB(+,-).\cite{theorydependenceAIM}

In the inset, the density vs.\ bond length is reported in the doubly log-scale. It shows that the bond lengths increase via the trend F$\ll$Cl<Br<I. 
Brammer \textit{et al.}\cite{Brammer2001} found the same trend except that they did not find a distinction between chlorine and bromine.
This distinction is clearly distinguishable here, although being smaller than F$\ll$Cl and Br<I.
The results follow a straight line in the log-log plot, implying a strong inverse power law relation. 
We performed a linear regression on these plots for the N-H$\cdots$X bonds for fluorine and iodine, the two  halogens at the opposite ends for bond lengths, giving the following equations:
\begin{equation}
\begin{split}
\rho_{BCP, Fluorine}^{N-H\cdots X}  &=0.36\cdot BL^{-4.1} \\
\rho_{BCP, Iodine}^{N-H\cdots X}  &=2.72\cdot BL^{-5.2}      
\end{split}
\end{equation}
where $BL$ is the bond length in Angstroms. 
The coefficients of determination, $R^2$ where 0.985 and 0.980 for Fluorine and Iodine respectively.
The larger prefactor in Iodine clearly illustrates the generally larger bond lengths in Iodine, while the larger power in Iodine, indicates that the density goes down faster when the bond lengths increase. Surprisingly, Gibbs \textit{et al.}\ could not find a physical interpretation for this observed power law.\cite{Gibbs1998}

All the gradients of the density at the BCPs are positive, indicating that all considered bonds are non-covalent and that electrostatic forces dominate.\cite{KUMAR2016, VARADWAJ2018} 
All the observed density values are between 0.00128 and 0.067 fully covering the values normally reported for weak interactions between 0.002 and 0.034. 
The density gradients are between 0.004 and 0.171 also covering the whole range of reported gradient values reported for weak values, i.e., between 0.024 and 0.139.\cite{KUMAR2016} 
We found some BCPs with densities and gradients that are even lower than the normally reported range for weak interactions; these BCPs are very weak hydrogen bonding interactions with iodide and bromine at distances around and larger than 4 \AA. 
\bibliography{citations}